\documentclass[10pt,nohyperref]{sigplanconf}

\toappear{Preprint produced \today\  for {\tt arXiv.org}.  Copyright \copyright\ 2015 Oracle.}

\usepackage{times}
\usepackage{graphicx} % more modern
\usepackage{subfigure} 

\usepackage{natbib}

\usepackage{algpseudocode}
\usepackage{algorithm}
\usepackage{algorithmicx}

\usepackage{latexsym}
\usepackage{color}
\usepackage{graphicx}

\multiply\textfloatsep by 1\divide\textfloatsep by 2
\multiply\dbltextfloatsep by 1\divide\dbltextfloatsep by 2

\algblockdefx{Begin}{End}{{\bf begin}}{{\bf end}}
\newcommand*\Assign[2]{\State #1 $\gets$ #2}
\newcommand*\LocalArray[1]{\State {\bf local array} #1}
\newcommand*\RegisterArray[1]{\State {\bf register array} #1}
\newcommand*\Bind[2]{\State {\bf let} #1 $\gets$ #2}
\newcommand*\PhantomBind[2]{\State{\setbox0\hbox{{\bf let} #1 $\gets$ (}\hbox to \wd0{\hss}#2}}
\newcommand*\Remark[1]{\State \(\triangleright\) #1}
\newcommand*\Var[1]{{\mathit{#1}}}
\newcommand*\Output{\;\mathbf{output}\;}

\newbox\proctwobox
\algblockdefx{ProcTwo}{EndProcedure}[2]{\global\setbox\proctwobox\hbox{{\bf procedure} {\sc #1}$\bigl($}\copy\proctwobox#2,}{{\bf end}}
\newcommand*\ProcTwoX[1]{\State\hbox{\hskip-\algorithmicindent\hskip\wd\proctwobox#1$\bigr)$}}

\algblockdefx{CodeChunk}{End}[1]{{\bf code chunk} $\langle${#1}$\rangle$:}{{\bf end}}
\newcommand*\UseCodeChunk[1]{\State $\langle${#1}$\rangle$}

\algrenewcommand\algorithmicindent{0.8em}

\renewcommand\topfraction{0.8}

\begin{document}

\conferenceinfo{no conference}{no place}
\copyrightyear{2015} 
\copyrightdata{no copyright data}
\doi{no doi}

\title{Using Butterfly-Patterned Partial Sums \\ to Optimize GPU Memory Accesses \\ for Drawing from Discrete Distributions}
\subtitle{}
\authorinfo{Guy L. Steele Jr.}{Oracle Labs}{guy.steele@oracle.com}
\authorinfo{Jean-Baptiste Tristan}{Oracle Labs}{jean.baptiste.tristan@oracle.com}

% \makeatletter
% \def \@maketitle {%
%  \begin{center}
%  \@settitlebanner
%  \let \thanks = \titlenote
%  \noindent \LARGE \bfseries \@titletext \par
%  %\vskip 6pt
%  %\noindent \Large \@subtitletext \par
%  \vskip 6pt
%    \noindent \@setauthor{9pc}{i}{\@false}\hspace{1.5pc}%
%              \@setauthor{9pc}{ii}{\@false}\hspace{1.5pc}%
%              \@setauthor{9pc}{iii}{\@false}\par
% \vspace{10pt plus 2pt}
%  \end{center}}
% \makeatother
\maketitle

\begin{abstract} 
We describe a technique for drawing values from discrete distributions,
such as sampling from the random variables of a mixture model,
that avoids computing a complete table of partial
sums of the relative probabilities.  A table of alternate
(``butterfly-patterned'') form is faster to compute, making
better use of coalesced memory accesses.  From this table,
complete partial sums are computed on the fly during a binary search.
Measurements using an {\sc nvidia} Titan Black {\sc gpu} show that for a sufficiently large number
of clusters or topics ($K > 200$), this technique alone more than doubles
the speed of a latent Dirichlet allocation ({\sc LDA}) application already highly tuned for GPU execution.
\end{abstract} 


\section{Overview}
\label{sec:overview}

The successful use of Graphics Processing Units ({\sc gpu}s) to train neural
networks is a great example of how machine learning can benefit from
such massively parallel architecture. Generative
probabilistic modeling \cite{Blei}
and associated inference methods (such as Monte Carlo
methods) can also benefit. Indeed,
authors such as Suchard~{\it et~al.}~\cite{suchard_understanding_2010} and
Lee~{\it et~al.}~\cite{lee_utility_journal_2010} have pointed out that many algorithms of
interest are embarrassingly parallel. However, the potential for massively
parallel computation is only the first step toward full
use of {\sc gpu} capacity.  One bottleneck that such embarrassingly parallel
algorithms run into is related to memory bandwidth; one must
design key probabilistic primitives with such constraints in mind.

We address the important case wherein parallel threads must
draw from distinct discrete distributions in a {\sc simd} fashion.
This can arise when implementing any mixture model, and Latent Dirichlet Allocation ({\sc lda}) models in particular,
which are probabilistic mixture models used to discover abstract ``topics'' in a collection of documents (a \emph{corpus}) \cite{blei_latent_2003}. This model can be fitted (or ``trained'') in an unsupervised fashion using sampling methods \cite[chapter 11]{Bishop}\cite{griffiths_finding_2004}. Each document is modeled as a distribution $\theta$ over topics, and each word in a document is assumed to be drawn from a distribution $\phi$ of words. Understanding the methods described in this paper does not require a deep understanding of sampling algorithms for {\sc lda}. What is important is that each word in a corpus is associated with a so-called ``latent'' random variable \cite[chapter 9]{Bishop}, usually referred to as $z$, that takes on one of $K$ integer values, indicating a topic to which the word belongs.  Broadly speaking, the iterative training process works by tentatively choosing a topic (that is, sampling the random latent variable $z$) for a given word using relative probabilities calculated from $\theta$ and $\phi$, then updating $\theta$ and $\phi$ accordingly.

In this paper, we focus on the step that draws new $z$ values from a finite domain,
given arrays of (typically floating-point) parameters $\theta$ and $\phi$,
typically using the following four-step process for each new $z$ value to be drawn:
\par\noindent{\bf 1.\quad}Use the parameters to construct a table of relative (unnormalized) probabilities for the possible choices for $z$,
where each relative probability is a product of some element of $\theta$ and some element of $\phi$.
\hfill\break{\bf 2.\quad}Normalize these table entries by dividing each entry by the sum of all entries.
\hfill\break{\bf 3.\quad}Let $u$ be chosen uniformly at random from the real interval $[0,1)$.
\hfill\break{\bf 4.\quad}Find the smallest index $j$ such that the sum of all table entries at or below index $j$ is larger than $u$.
\par\noindent In practice, this sequence of steps may be optimized
by doing a bit of algebra and using a binary search:
\par\noindent{\bf 1.\quad}Use the parameters to construct a table of relative probabilities for the possible choices for the $z$ value.
\hfill\break{\bf 2.\quad}Replace this table with its sum-prefix: each entry is replaced by the sum of itself and all earlier entries.  The last entry therefore becomes the sum of all the original entries.
\hfill\break{\bf 3.\quad}Let $u$ be chosen uniformly at random from the real interval $[0,1)$.
\hfill\break{\bf 4.\quad}Use a binary search to find the smallest index $j$ such that the entry at index $j$ is larger than $u$ times the last entry.

When this algorithm is implemented in parallel on a {\sc simd} {\sc gpu}, an obvious approach is to assign the computation of each $z$ value to a separate thread.
However, when the threads fetch their respective $\phi$ values, the values to be fetched will likely reside at unrelated locations in memory, resulting in poor memory-fetch performance.  A standard trick is to have all the threads in a warp cooperate with each other (compare, for example, the storage of floating-point numbers as ``slicewise'' rather than ``fieldwise'' in the architecture of the Connection Machine Model CM-2, so that 32 1-bit processors cooperate on each of 32 clock cycles to fetch and store an entire 32-bit floating-point value that logically belongs to just one of the 32 processors~\cite{Johnsson-1989}).  Suppose there are $W$ threads in a warp (a typical value is $W=32$), each working on a different document (and therefore using different discrete probability distributions), and suppose that at a certain step of the algorithm each thread needs to fetch $W$ different $\phi$ values.  The array of $\phi$ values can be organized so that the $W$ different $\phi$ values needed by any one thread are stored in consecutive memory locations.  The trick is that each thread performs $W$ fetches from the $\phi$ table, as before, but instead of each thread fetching its own $\phi$ values in all cases, on cycle $k$ all the threads work together to fetch the $W$ different $\phi$ values needed by thread $k$.  As a result, on each memory cycle all $W$ values being fetched are in a contiguous region of memory,
allowing improved memory-fetch performance.

It is then necessary for the threads to exchange information among themselves so that the rest of the algorithm may be carried out, including the summation arithmetic.

The binary search does not access all entries in the prefix-sum table (in fact, for a table of size $N$ it examines only about $\log_2 N$ entries).  Therefore it is not necessary to compute all entries of the prefix-sum table.  We present
an alternate technique that computes
a ``butterfly-patterned'' partial-sums table, using less computational and communication effort; a modified binary search then uses this table to compute, on the fly, entries that would have been in the original complete prefix-sum table.  This requires more work per table entry during the binary search, but because the search examines only a few table entries, the result is a dramatic net reduction in execution time.  This technique may be effective for collapsed LDA Gibbs samplers~\cite{yan_parallel_2009,lu_accelerating_2013,Augur}
as well as uncollapsed samplers, and may also be useful for GPU implementations of other algorithms \cite{zhao_same_2014}
whose inner loops sample from discrete distributions.

\section{Basic Algorithm}

For a Latent Dirichlet Allocation model of, for example, a set of documents to which we want to assign topics probablistically
using Gibbs sampling, let $M$ be the number of documents,
$K$ be the number of topics, and $V$ be the size of the vocabulary, which is a set of distinct words.
Each document is a bag of words, each of which belongs to the vocabulary;
any given word can appear in any number of documents, and may appear any number of times in any single document.
The documents may be of different lengths.

We are interested in the phase of an uncollapsed Gibbs sampler that draws new $z$ values, given $\theta$ and $\phi$ distributions.
Because no $z$ value directly depends on any other $z$ value in this formulation, new $z$ values may all be computed independently
(and therefore in parallel to any extent desired).

We assume that we are given an $M \times K$ matrix $\theta$ and a $V \times K$ matrix $\phi$; the elements of these matrices are
non-negative numbers, typically represented as floating-point values.
Row $m$ of $\theta$ (that is, $\theta[m, \cdot]$) is the (currently assumed) distribution of topics for document $m$, that is,
the relative probabilities (weights) for each of the $K$ possible topics to which the document might be assigned.
Note that columns of $\theta$ are \emph{not} to be considered as distributions.
Similarly, column $k$ of $\phi$ (that is, $\phi[\cdot,k$) is the (currently assumed) distribution of words for topic $k$,
that is, the weights with which the $V$ possible words in the vocabulary are associated with the topic.
Note that rows of $\phi$ are \emph{not} to be considered as distributions.
We organize $\theta$ as rows and $\phi$ as columns for engineering reasons:
we want the $K$ entries obtained by ranging over all possible topics to be contiguous in memory
so as to take advantage of memory cache structure.

We also assume that we are given (i)~a length-$M$
vector of nonnegative integers $N$ such that $N[m]$ is the length of document $m$, and (ii)~an $M \times N$ ragged array $w$,
by which we mean that for $0 \leq m < M$, $w[m]$ is a vector of length $N[m]$.  (We use zero-based indexing throughout this document.)
Each element of $w$ is less than $V$ and may therefore be used as a first index for $\phi$.

\begin{algorithm}[t]
\caption{Drawing new $z$ values}\label{alg:basicdraw}
\begin{algorithmic}[1]
\ProcTwo{DrawZ}{$N[M], \theta[M, K], \phi[V, K]$}\ProcTwoX{$w[M][N]; \Output z[M, N]$}
   \LocalArray{$a[M][N][K], p[M][N][K]$}
   \ForAll{$0 \leq m < M$}
      \ForAll{$0 \leq i < N[m]$}
         \Remark{Compute $\theta$-$\phi$ products}
         \ForAll{$0 \leq k < K$}   \label{z1prodstart}
            \Assign{$a[m][i][k]$}{$\theta[m, k] \times \phi[w[m][i], k]$}
         \EndFor   \label{z1prodend}
         \Remark{Compute partials sums of the products}
         \Bind{$\Var{sum}$}{$0.0$}  \label{z1sumstart}
         \For{$k$ from $0$ through $K-1$}
            \Assign{$\Var{sum}$}{$\Var{sum} + a[m][i][k]$}
            \Assign{$p[m][i][k]$}{$\Var{sum}$}
         \EndFor \label{z1sumend}
         \Bind{$j$}{$0$}
         \UseCodeChunk{search the table of partial sums}
         \Assign{$z[m, i]$}{$j$} \label{z1searchend}
     \EndFor
   \EndFor
\EndProcedure
\end{algorithmic}
\end{algorithm}

Our goal, given $K$, $M$, $V$, $N$, $\phi$, $\theta$, and $w$ and
assuming the use of a temporary $M \times N \times K$ ragged work array $a$ (which we will later optimize away), is to compute
all the elements for an $M \times N$ ragged array $z$ as follows:
For all $m$ such that $0 \leq m < M$ and for all $i$ such that $0 \leq i < N[m]$, do two things:
first, for all $k$ such that $0 \leq k < K$, let $a[m][i][k] = \theta[m, k] \times \phi[w[m, i], k]$;
second, let $z[m, i]$ be a nonnegative integer less than $K$, chosen randomly in such a way that the probability of choosing
the value $k'$ is $a[m][i][k']/\sigma$ where
$\sigma = \sum\limits_{0 \leq k < K} a[m][i][k]$.
Thus, $a[m][i][k']$ is a relative (unnormalized) probability, and $a[m][i][k']/\sigma$ is an absolute (normalized) probability.

Algorithm~\ref{alg:basicdraw} is a basic implementation of this process.
We remark that a ``{\bf let}'' statement creates a local binding of a scalar (single-valued) variable and gives it a value,
that a ``{\bf local array}'' declaration creates a local binding of an array variable
(containing an element value for each indexable position in the array), and that distinct iterations of a ``{\bf for}'' or ``{\bf for all}'' construct are understood to create
distinct and independent instantiations of such local variables for each iteration.  The iterations of ``{\bf for} \ldots\ from \ldots\ through \ldots''
are executed sequentially in a specific order; but the iterations of a ``{\bf for all}'' construct are intended to be computationally independent
and therefore may be executed in any order, or in parallel, or in any sequential-parallel combination.
We use angle brackets to indicate the use of a ``code chunk'' that is defined as a separate algorithm;
such a use indicates that the definition of the code chunk should be inserted at the use site, as if it were a C macro,
but surrounded by {\bf begin} and {\bf end} (this is a programming-language
technicality that ensures that the scope of any variable declared within the code chunk is confined to that code chunk).

The computation of the $\theta$-$\phi$ products (lines~\hbox{\ref{z1prodstart}--\ref{z1prodend}} of Algorithm~\ref{alg:basicdraw}) is straightforward.
The computation of partial sums (lines~\hbox{\ref{z1sumstart}--\ref{z1sumend}}) is sequential; the variable $\Var{sum}$
accumulates the products, and successive values of $\Var{sum}$ are stored into the array $p$.
A random integer is chosen for $z[m,i]$ by choosing a random value uniformly from the range $[0,0, 1.0)$,
scaling it by the final value of $sum$ (which has the same algorithmic effect as dividing each $p[m][i][k]$
by that value, for all $0 \leq k < K$, to turn it into an absolute probability), and then searching the subarray $p[m][i]$
to find the smallest entry that is larger than the scaled value (and if there are several such entries, all equal,
then the one with the smallest index is chosen); the index $j$ of that entry is used as the desired
randomly chosen integer.   A simple linear search (Algorithm~\ref{alg:basiclinearsearch}) can do the job.
But because all elements of $\theta$ and $\phi$ are nonnegative,
the products in $a$ are also nonnegative, and so each subarray $p[m][i]$ is monotonically nondecreasing;
that is, for all $0 \leq m < M$, $0 \leq i < N[m]$, and $0 < k < K$, we have $p[m][i][k-1] \leq p[m][i][k]$.
Therefore a binary search (Algorithm~\ref{alg:basicbinarysearch}) can be used instead, which is faster,
on average, for $K$ sufficiently large
%(\citeauthor{KNUTH-VOLUME-3}, \citeyear{KNUTH-VOLUME-3}, exercise 6.2.1-5).
\cite[exercise 6.2.1-5]{KNUTH-VOLUME-3}.

\begin{algorithm}[t]
\caption{Simple linear search}\label{alg:basiclinearsearch}
\begin{algorithmic}[1]
\CodeChunk{search the table of partial sums}
         \Bind{$u$}{random value chosen from $[0.0,1.0)$}
         \Bind{$\Var{stop}$}{$\Var{sum} \times u$}
        \While{$j < K-1$ and $\Var{stop} \geq p[m][i][j]$}
            \Assign{$j$}{$j + 1$}
         \EndWhile
\End
\end{algorithmic}
\end{algorithm}

\begin{algorithm}[t]
\caption{Simple binary search}\label{alg:basicbinarysearch}
\begin{algorithmic}[1]
\CodeChunk{search the table of partial sums}
         \Bind{$u$}{random value chosen from $[0.0,1.0)$}
         \Bind{$\Var{stop}$}{$\Var{sum} \times u$}
            \Bind{$k$}{$K-1$}
            \While{$j < k$}
               \Bind{$\Var{mid}$}{$\displaystyle \left\lfloor \frac{j + k}{2} \right\rfloor$}
               \If{$\Var{stop} < p[m][i][\Var{mid}]$}
                  \Assign{$k$}{$\Var{mid}$}
               \Else
                  \Assign{$j$}{$\Var{mid} + 1$}
               \EndIf
            \EndWhile
\End
\end{algorithmic}
\end{algorithm}

\section{Blocking and Transposition}

Anticipating certain characteristics of the hardware,
we make some commitments as to how the algorithm will be executed.  We assume that
arrays are laid out in row-major order (as they are when using C or {\sc cuda}).
Let $W$ be a
machine-dependent constant (typically 16 or 32, but for now we do not require that $W$ be a power of 2).
For purposes of illustration we assume $W=8$ and also $K=19$.
We divide the documents into groups of size $W$ and assume that $M$ is an exact multiple of $W$.
(In an overall application, the set of documents can be padded with empty documents
so as to make $M$ be an exact multiple of $W$ without affecting the overall behavior of the algorithm on the ``real'' documents.)
We split the outermost loop of Algorithm~\ref{alg:basicdraw} (with index variable $m$) into two nested loops with
index variables $q$ and $r$, from which the equivalent value for $m$ is then computed.
We commit to making the loop with index variable $i$ sequential,
to treating the iterations of the loop on $q$ as independent (and therefore possibly parallel),
and to treating the iterations of the loop on $r$ as executed by a {\sc simd} ``thread warp'' of size $W$, that is,
parallel and implicitly lock-step synchronized.
As a result, we view each of the $M$ documents as being processed by a separate thread.
A benefit of making the loop on $i$ sequential is that the array $p$ can be made two-dimensional and non-ragged,
having size $M \times K$.  We fuse the loop that computes $\theta$-$\phi$ products
with the loop that computes partial sums; this eliminates the need for the array $a$,
but instead (for reasons explained below) we retain $a$ as a two-dimensional, non-ragged array of size $M \times W$
that is used only when $K \geq W$.  Within the loop controlling index variable $q$ we declare a local work array $c_{\mathrm{warp}}$
of size $W \times W$ that will be used to exchange information by the $W$ threads within a warp; our eventual intent
is that this array will reside in {\sc gpu} registers.
We cache values from the array $\theta$ in a per-thread array $\theta_{\mathrm{local}}$ of length $K$,
anticipating that such cached values will reside in a faster memory and
be used repeatedly by the loop on $i$.

\begin{algorithm}[t]
\caption{Drawing $z$ values (transposed access)}\label{alg:transposedraw}
\begin{algorithmic}[1]
\ProcTwo{DrawZ}{$N[M], \theta[M, K], \phi[V, K]$}\ProcTwoX{$w[M][N]; \Output z[M, N]$}
  \LocalArray{$p[M, K], a[M,W]$}
   \ForAll{$0 \leq q < M/W$}
      \LocalArray{$c_{\mathrm{warp}}[W,W]$}
      \For{\hbox{\small\bf SIMD} $0 \leq r < W$}
         \Bind{$m$}{$q \times W + r$}
         \LocalArray{$\theta_{\mathrm{local}}[K]$}
         \UseCodeChunk{cache $\theta$ values into $\theta_{\mathrm{local}}$}
         \Bind{$i_{\mathrm{master}}$}{$0$}    \label{z6iloopstart}
          \While{$\Var{any}(i_{\mathrm{master}} < N[m])$}
            \Bind{$i$}{$\Var{min}(i_{\mathrm{master}}, N[m]-1)$}
            \UseCodeChunk{compute partial sums of $\theta$-$\phi$ products}
            \Bind{$j$}{$0$}
            \UseCodeChunk{search the table of partial sums}
            \Assign{$z[m, i]$}{$j$}
            \Assign{$i_{\mathrm{master}}$}{$i_{\mathrm{master}} + 1$}
         \EndWhile     \label{z6iloopend}
   \EndFor
   \EndFor
\EndProcedure
\end{algorithmic}
\end{algorithm}

There is, however, a subtle problem with the loop controlling index variable $i$: the upper bound $N[m]$ for the loop
variable may be different for different threads.  As a result, in the last iterations it may be that some threads
have ``gone to sleep'' because they reached their upper loop bound earlier than other threads in the warp.
This is undesirable because, as we shall see, we rely on all threads ``staying awake'' so that they can assist
each other.  Therefore, we rewrite the loop control to use a ``master index'' idiom and exploit the trick of allowing a thread
to perform its last iteration (with $i=N[m]-1$) multiple times, which doesn't work for many algorithms
but is acceptable for {\sc lda} Gibbs.

The result of all these code transformations is Algorithm~\ref{alg:transposedraw},
which makes use of three code chunks: Algorithms~\ref{alg:transposecache}, \ref{alg:transposesum}, and either
Algorithm~\ref{alg:basiclinearsearch} or~\ref{alg:basicbinarysearch}.
Algorithms~\ref{alg:transposecache} and~\ref{alg:transposesum}, besides using {\sc simd} thread warps of size $W$
to process documents in groups of size $W$, also process topics in blocks of size $W$.  This allows the innermost loops to process ``little'' arrays
of size $W \times W$.  If $K$ (the number of topics) is not a multiple of $W$, then there will be a \emph{remnant} of size $K \bmod W$.
To make looping code slightly simpler, we put the remnant at the \emph{front} of each array, rather than at the end.
For $W=8$ and $K=19$, topics 0, 1, and 2 form the remnant; topics 3--10 form a block of length 8;
and topics 11--18 form a second block.  This organization of arrays into blocks
allows reduction of the cost of accessing data in main memory by performing \emph{transposed accesses}.

\begin{algorithm}[t]
\caption{Caching $\theta$ values (transposed access)}\label{alg:transposecache}
\begin{algorithmic}[1]
\CodeChunk{cache $\theta$ values into $\theta_{\mathrm{local}}$}
          \Bind{$j$}{$0$}
            \While{$j < (K \bmod W)$}   \label{z6cacheremnantloopstart}
               \Assign{$\theta_{\mathrm{local}}[j]$}{$\theta[m, j]$}
               \Assign{$j$}{$j + 1$}
            \EndWhile    \label{z6cacheremnantloopend}
            \While{$j < K$}
               \For{$k$ from $0$ through $W-1$}   \label{z6cacheblockloopstart}
                  \Remark{Next line uses transposed access to $\theta$}
                  \Assign{$\theta_{\mathrm{local}}[j+k]$}{$\theta[q \times W + k, j+r]$}     \label{z6thetaaccess}
               \EndFor    \label{z6cacheblockloopend}
              \Assign{$j$}{$j + W$}
            \EndWhile
\End
\end{algorithmic}
\end{algorithm}

The simplest use of transposed memory access occurs in Algorithm~\ref{alg:transposecache}.
For every document, this algorithm fetches a $\theta$ value for every topic.
The topics are regarded as divided into a leading remnant (if any) and then a sequence
of blocks of length $W$.  The {\bf while} loop on lines~\hbox{\ref{z6cacheremnantloopstart}--\ref{z6cacheremnantloopend}}
handles the remnant, and then the following {\bf while} loop processes successive blocks.
On line~\ref{z6thetaaccess} within the inner loop,
note that the reference is to $\theta[q \times W + k, j+r]$ rather than the expected $\theta[q \times W + r, j+k]$
(which would be the same as $\theta[m, j+k]$ because $m=q \times W + r$).  The result is that when
the $W$ threads of a {\sc simd} warp execute this code and all access $\theta$ simultaneously, they access
$W$ consecutive memory locations, which can typically be fetched by a hardware memory controller
much more efficiently than $W$ memory locations separated by stride $K$.
Another way to think about it is that on any given single iteration of the loop on
lines~\hbox{\ref{z6cacheblockloopstart}--\ref{z6cacheblockloopend}}
(which overall is designed to fetch one $W\times W$ block of $\theta$ values)
instead of every thread in the warp fetching its $k$th value from the $\theta$ array,
all the threads work together to fetch all $W$ values that are needed by thread $k$ of the warp.
Each thread then stores what it has fetched into its local copy of the array $\theta_{\mathrm{local}}$.
The result is that each thread doesn't really have all the $\theta$ data it needs to process its
document; pictorially, data in each $W \times W$ block has been \emph{transposed}.
The real story, however, is that memory for local arrays in a {\sc gpu} is not stored in the same way as memory for global arrays.
The global array $\theta$ is laid out in row-major order in main memory like this:

\noindent
\hbox to \linewidth{\relax
\hsize=0.4\linewidth
\vbox{\unitlength=1 in \divide\unitlength by 16
\thinlines
\def\SQX{\vbox to 0pt{\vss\hbox to 0pt{\hss\vrule width 0.6\unitlength height 0.6\unitlength\hss}\vss}}\noindent
\begin{picture}(19,8)
\color[gray]{0}
\put(0,8){\line(1,0){19}}
\put(0,0){\line(1,0){19}}
\put(19,0){\line(0,1){8}}
\put(11,0){\line(0,1){8}}
\put(3,0){\line(0,1){8}}
\put(0,0){\line(0,1){8}}
\color[gray]{0.00}
\put(0.5,7.5){\SQX}
\put(1.5,7.5){\SQX}
\put(2.5,7.5){\SQX}
\put(3.5,7.5){\SQX}
\put(4.5,7.5){\SQX}
\put(5.5,7.5){\SQX}
\put(6.5,7.5){\SQX}
\put(7.5,7.5){\SQX}
\put(8.5,7.5){\SQX}
\put(9.5,7.5){\SQX}
\put(10.5,7.5){\SQX}
\put(11.5,7.5){\SQX}
\put(12.5,7.5){\SQX}
\put(13.5,7.5){\SQX}
\put(14.5,7.5){\SQX}
\put(15.5,7.5){\SQX}
\put(16.5,7.5){\SQX}
\put(17.5,7.5){\SQX}
\put(18.5,7.5){\SQX}
\color[gray]{0.12}
\put(0.5,6.5){\SQX}
\put(1.5,6.5){\SQX}
\put(2.5,6.5){\SQX}
\put(3.5,6.5){\SQX}
\put(4.5,6.5){\SQX}
\put(5.5,6.5){\SQX}
\put(6.5,6.5){\SQX}
\put(7.5,6.5){\SQX}
\put(8.5,6.5){\SQX}
\put(9.5,6.5){\SQX}
\put(10.5,6.5){\SQX}
\put(11.5,6.5){\SQX}
\put(12.5,6.5){\SQX}
\put(13.5,6.5){\SQX}
\put(14.5,6.5){\SQX}
\put(15.5,6.5){\SQX}
\put(16.5,6.5){\SQX}
\put(17.5,6.5){\SQX}
\put(18.5,6.5){\SQX}
\color[gray]{0.24}
\put(0.5,5.5){\SQX}
\put(1.5,5.5){\SQX}
\put(2.5,5.5){\SQX}
\put(3.5,5.5){\SQX}
\put(4.5,5.5){\SQX}
\put(5.5,5.5){\SQX}
\put(6.5,5.5){\SQX}
\put(7.5,5.5){\SQX}
\put(8.5,5.5){\SQX}
\put(9.5,5.5){\SQX}
\put(10.5,5.5){\SQX}
\put(11.5,5.5){\SQX}
\put(12.5,5.5){\SQX}
\put(13.5,5.5){\SQX}
\put(14.5,5.5){\SQX}
\put(15.5,5.5){\SQX}
\put(16.5,5.5){\SQX}
\put(17.5,5.5){\SQX}
\put(18.5,5.5){\SQX}
\color[gray]{0.36}
\put(0.5,4.5){\SQX}
\put(1.5,4.5){\SQX}
\put(2.5,4.5){\SQX}
\put(3.5,4.5){\SQX}
\put(4.5,4.5){\SQX}
\put(5.5,4.5){\SQX}
\put(6.5,4.5){\SQX}
\put(7.5,4.5){\SQX}
\put(8.5,4.5){\SQX}
\put(9.5,4.5){\SQX}
\put(10.5,4.5){\SQX}
\put(11.5,4.5){\SQX}
\put(12.5,4.5){\SQX}
\put(13.5,4.5){\SQX}
\put(14.5,4.5){\SQX}
\put(15.5,4.5){\SQX}
\put(16.5,4.5){\SQX}
\put(17.5,4.5){\SQX}
\put(18.5,4.5){\SQX}
\color[gray]{0.48}
\put(0.5,3.5){\SQX}
\put(1.5,3.5){\SQX}
\put(2.5,3.5){\SQX}
\put(3.5,3.5){\SQX}
\put(4.5,3.5){\SQX}
\put(5.5,3.5){\SQX}
\put(6.5,3.5){\SQX}
\put(7.5,3.5){\SQX}
\put(8.5,3.5){\SQX}
\put(9.5,3.5){\SQX}
\put(10.5,3.5){\SQX}
\put(11.5,3.5){\SQX}
\put(12.5,3.5){\SQX}
\put(13.5,3.5){\SQX}
\put(14.5,3.5){\SQX}
\put(15.5,3.5){\SQX}
\put(16.5,3.5){\SQX}
\put(17.5,3.5){\SQX}
\put(18.5,3.5){\SQX}
\color[gray]{0.60}
\put(0.5,2.5){\SQX}
\put(1.5,2.5){\SQX}
\put(2.5,2.5){\SQX}
\put(3.5,2.5){\SQX}
\put(4.5,2.5){\SQX}
\put(5.5,2.5){\SQX}
\put(6.5,2.5){\SQX}
\put(7.5,2.5){\SQX}
\put(8.5,2.5){\SQX}
\put(9.5,2.5){\SQX}
\put(10.5,2.5){\SQX}
\put(11.5,2.5){\SQX}
\put(12.5,2.5){\SQX}
\put(13.5,2.5){\SQX}
\put(14.5,2.5){\SQX}
\put(15.5,2.5){\SQX}
\put(16.5,2.5){\SQX}
\put(17.5,2.5){\SQX}
\put(18.5,2.5){\SQX}
\color[gray]{0.72}
\put(0.5,1.5){\SQX}
\put(1.5,1.5){\SQX}
\put(2.5,1.5){\SQX}
\put(3.5,1.5){\SQX}
\put(4.5,1.5){\SQX}
\put(5.5,1.5){\SQX}
\put(6.5,1.5){\SQX}
\put(7.5,1.5){\SQX}
\put(8.5,1.5){\SQX}
\put(9.5,1.5){\SQX}
\put(10.5,1.5){\SQX}
\put(11.5,1.5){\SQX}
\put(12.5,1.5){\SQX}
\put(13.5,1.5){\SQX}
\put(14.5,1.5){\SQX}
\put(15.5,1.5){\SQX}
\put(16.5,1.5){\SQX}
\put(17.5,1.5){\SQX}
\put(18.5,1.5){\SQX}
\color[gray]{0.84}
\put(0.5,0.5){\SQX}
\put(1.5,0.5){\SQX}
\put(2.5,0.5){\SQX}
\put(3.5,0.5){\SQX}
\put(4.5,0.5){\SQX}
\put(5.5,0.5){\SQX}
\put(6.5,0.5){\SQX}
\put(7.5,0.5){\SQX}
\put(8.5,0.5){\SQX}
\put(9.5,0.5){\SQX}
\put(10.5,0.5){\SQX}
\put(11.5,0.5){\SQX}
\put(12.5,0.5){\SQX}
\put(13.5,0.5){\SQX}
\put(14.5,0.5){\SQX}
\put(15.5,0.5){\SQX}
\put(16.5,0.5){\SQX}
\put(17.5,0.5){\SQX}
\put(18.5,0.5){\SQX}
\end{picture}\vskip5pt
}\hfil
{\hsize=0.6\linewidth
\vbox{\noindent\strut
where each row corresponds to a document and each column to a topic; but each of the local arrays is laid out like this:\strut}}}

\unskip\unskip
\hbox to \linewidth{\relax
\hsize=0.2\linewidth
\vbox{\unitlength=1 in \divide\unitlength by 16
\thinlines
\def\SQX{\vbox to 0pt{\vss\hbox to 0pt{\hss\vrule width 0.6\unitlength height 0.6\unitlength\hss}\vss}}\noindent
\begin{picture}(8,19)
\color[gray]{0}
\put(0,0){\line(0,1){19}}
\put(8,0){\line(0,1){19}}
\put(0,0){\line(1,0){8}}
\put(0,8){\line(1,0){8}}
\put(0,16){\line(1,0){8}}
\put(0,19){\line(1,0){8}}
\color[gray]{0.00}
\put(0.5,18.5){\SQX}
\put(0.5,17.5){\SQX}
\put(0.5,16.5){\SQX}
\color[gray]{0.00}\put(0.5,15.5){\SQX}
\color[gray]{0.12}\put(0.5,14.5){\SQX}
\color[gray]{0.24}\put(0.5,13.5){\SQX}
\color[gray]{0.36}\put(0.5,12.5){\SQX}
\color[gray]{0.48}\put(0.5,11.5){\SQX}
\color[gray]{0.60}\put(0.5,10.5){\SQX}
\color[gray]{0.72}\put(0.5,9.5){\SQX}
\color[gray]{0.84}\put(0.5,8.5){\SQX}
\color[gray]{0.00}\put(0.5,7.5){\SQX}
\color[gray]{0.12}\put(0.5,6.5){\SQX}
\color[gray]{0.24}\put(0.5,5.5){\SQX}
\color[gray]{0.36}\put(0.5,4.5){\SQX}
\color[gray]{0.48}\put(0.5,3.5){\SQX}
\color[gray]{0.60}\put(0.5,2.5){\SQX}
\color[gray]{0.72}\put(0.5,1.5){\SQX}
\color[gray]{0.84}\put(0.5,0.5){\SQX}
\color[gray]{0.12}
\put(1.5,18.5){\SQX}
\put(1.5,17.5){\SQX}
\put(1.5,16.5){\SQX}
\color[gray]{0.00}\put(1.5,15.5){\SQX}
\color[gray]{0.12}\put(1.5,14.5){\SQX}
\color[gray]{0.24}\put(1.5,13.5){\SQX}
\color[gray]{0.36}\put(1.5,12.5){\SQX}
\color[gray]{0.48}\put(1.5,11.5){\SQX}
\color[gray]{0.60}\put(1.5,10.5){\SQX}
\color[gray]{0.72}\put(1.5,9.5){\SQX}
\color[gray]{0.84}\put(1.5,8.5){\SQX}
\color[gray]{0.00}\put(1.5,7.5){\SQX}
\color[gray]{0.12}\put(1.5,6.5){\SQX}
\color[gray]{0.24}\put(1.5,5.5){\SQX}
\color[gray]{0.36}\put(1.5,4.5){\SQX}
\color[gray]{0.48}\put(1.5,3.5){\SQX}
\color[gray]{0.60}\put(1.5,2.5){\SQX}
\color[gray]{0.72}\put(1.5,1.5){\SQX}
\color[gray]{0.84}\put(1.5,0.5){\SQX}
\color[gray]{0.24}
\put(2.5,18.5){\SQX}
\put(2.5,17.5){\SQX}
\put(2.5,16.5){\SQX}
\color[gray]{0.00}\put(2.5,15.5){\SQX}
\color[gray]{0.12}\put(2.5,14.5){\SQX}
\color[gray]{0.24}\put(2.5,13.5){\SQX}
\color[gray]{0.36}\put(2.5,12.5){\SQX}
\color[gray]{0.48}\put(2.5,11.5){\SQX}
\color[gray]{0.60}\put(2.5,10.5){\SQX}
\color[gray]{0.72}\put(2.5,9.5){\SQX}
\color[gray]{0.84}\put(2.5,8.5){\SQX}
\color[gray]{0.00}\put(2.5,7.5){\SQX}
\color[gray]{0.12}\put(2.5,6.5){\SQX}
\color[gray]{0.24}\put(2.5,5.5){\SQX}
\color[gray]{0.36}\put(2.5,4.5){\SQX}
\color[gray]{0.48}\put(2.5,3.5){\SQX}
\color[gray]{0.60}\put(2.5,2.5){\SQX}
\color[gray]{0.72}\put(2.5,1.5){\SQX}
\color[gray]{0.84}\put(2.5,0.5){\SQX}
\color[gray]{0.36}
\put(3.5,18.5){\SQX}
\put(3.5,17.5){\SQX}
\put(3.5,16.5){\SQX}
\color[gray]{0.00}\put(3.5,15.5){\SQX}
\color[gray]{0.12}\put(3.5,14.5){\SQX}
\color[gray]{0.24}\put(3.5,13.5){\SQX}
\color[gray]{0.36}\put(3.5,12.5){\SQX}
\color[gray]{0.48}\put(3.5,11.5){\SQX}
\color[gray]{0.60}\put(3.5,10.5){\SQX}
\color[gray]{0.72}\put(3.5,9.5){\SQX}
\color[gray]{0.84}\put(3.5,8.5){\SQX}
\color[gray]{0.00}\put(3.5,7.5){\SQX}
\color[gray]{0.12}\put(3.5,6.5){\SQX}
\color[gray]{0.24}\put(3.5,5.5){\SQX}
\color[gray]{0.36}\put(3.5,4.5){\SQX}
\color[gray]{0.48}\put(3.5,3.5){\SQX}
\color[gray]{0.60}\put(3.5,2.5){\SQX}
\color[gray]{0.72}\put(3.5,1.5){\SQX}
\color[gray]{0.84}\put(3.5,0.5){\SQX}
\color[gray]{0.48}
\put(4.5,18.5){\SQX}
\put(4.5,17.5){\SQX}
\put(4.5,16.5){\SQX}
\color[gray]{0.00}\put(4.5,15.5){\SQX}
\color[gray]{0.12}\put(4.5,14.5){\SQX}
\color[gray]{0.24}\put(4.5,13.5){\SQX}
\color[gray]{0.36}\put(4.5,12.5){\SQX}
\color[gray]{0.48}\put(4.5,11.5){\SQX}
\color[gray]{0.60}\put(4.5,10.5){\SQX}
\color[gray]{0.72}\put(4.5,9.5){\SQX}
\color[gray]{0.84}\put(4.5,8.5){\SQX}
\color[gray]{0.00}\put(4.5,7.5){\SQX}
\color[gray]{0.12}\put(4.5,6.5){\SQX}
\color[gray]{0.24}\put(4.5,5.5){\SQX}
\color[gray]{0.36}\put(4.5,4.5){\SQX}
\color[gray]{0.48}\put(4.5,3.5){\SQX}
\color[gray]{0.60}\put(4.5,2.5){\SQX}
\color[gray]{0.72}\put(4.5,1.5){\SQX}
<\color[gray]{0.84}\put(4.5,0.5){\SQX}
\color[gray]{0.60}
\put(5.5,18.5){\SQX}
\put(5.5,17.5){\SQX}
\put(5.5,16.5){\SQX}
\color[gray]{0.00}\put(5.5,15.5){\SQX}
\color[gray]{0.12}\put(5.5,14.5){\SQX}
\color[gray]{0.24}\put(5.5,13.5){\SQX}
\color[gray]{0.36}\put(5.5,12.5){\SQX}
\color[gray]{0.48}\put(5.5,11.5){\SQX}
\color[gray]{0.60}\put(5.5,10.5){\SQX}
\color[gray]{0.72}\put(5.5,9.5){\SQX}
\color[gray]{0.84}\put(5.5,8.5){\SQX}
\color[gray]{0.00}\put(5.5,7.5){\SQX}
\color[gray]{0.12}\put(5.5,6.5){\SQX}
\color[gray]{0.24}\put(5.5,5.5){\SQX}
\color[gray]{0.36}\put(5.5,4.5){\SQX}
\color[gray]{0.48}\put(5.5,3.5){\SQX}
\color[gray]{0.60}\put(5.5,2.5){\SQX}
\color[gray]{0.72}\put(5.5,1.5){\SQX}
\color[gray]{0.84}\put(5.5,0.5){\SQX}
\color[gray]{0.72}
\put(6.5,18.5){\SQX}
\put(6.5,17.5){\SQX}
\put(6.5,16.5){\SQX}
\color[gray]{0.00}\put(6.5,15.5){\SQX}
\color[gray]{0.12}\put(6.5,14.5){\SQX}
\color[gray]{0.24}\put(6.5,13.5){\SQX}
\color[gray]{0.36}\put(6.5,12.5){\SQX}
\color[gray]{0.48}\put(6.5,11.5){\SQX}
\color[gray]{0.60}\put(6.5,10.5){\SQX}
\color[gray]{0.72}\put(6.5,9.5){\SQX}
\color[gray]{0.84}\put(6.5,8.5){\SQX}
\color[gray]{0.00}\put(6.5,7.5){\SQX}
\color[gray]{0.12}\put(6.5,6.5){\SQX}
\color[gray]{0.24}\put(6.5,5.5){\SQX}
\color[gray]{0.36}\put(6.5,4.5){\SQX}
\color[gray]{0.48}\put(6.5,3.5){\SQX}
\color[gray]{0.60}\put(6.5,2.5){\SQX}
\color[gray]{0.72}\put(6.5,1.5){\SQX}
\color[gray]{0.84}\put(6.5,0.5){\SQX}
\color[gray]{0.84}
\put(7.5,18.5){\SQX}
\put(7.5,17.5){\SQX}
\put(7.5,16.5){\SQX}
\color[gray]{0.00}\put(7.5,15.5){\SQX}
\color[gray]{0.12}\put(7.5,14.5){\SQX}
\color[gray]{0.24}\put(7.5,13.5){\SQX}
\color[gray]{0.36}\put(7.5,12.5){\SQX}
\color[gray]{0.48}\put(7.5,11.5){\SQX}
\color[gray]{0.60}\put(7.5,10.5){\SQX}
\color[gray]{0.72}\put(7.5,9.5){\SQX}
\color[gray]{0.84}\put(7.5,8.5){\SQX}
\color[gray]{0.00}\put(7.5,7.5){\SQX}
\color[gray]{0.12}\put(7.5,6.5){\SQX}
\color[gray]{0.24}\put(7.5,5.5){\SQX}
\color[gray]{0.36}\put(7.5,4.5){\SQX}
\color[gray]{0.48}\put(7.5,3.5){\SQX}
\color[gray]{0.60}\put(7.5,2.5){\SQX}
\color[gray]{0.72}\put(7.5,1.5){\SQX}
\color[gray]{0.84}\put(7.5,0.5){\SQX}
\end{picture}\vskip3pt
}\hfil
{\hsize=0.8\linewidth
\vbox{\noindent\strut\parfillskip=0pt
where each row corresponds to an array index and each column corresponds to a thread.  In each diagram,
the locations in a row are contiguous in memory; it is really the fact that we choose
to index $\theta_{\mathrm{local}}$ by topic and to assign each document to a thread
that causes ``transposition'' to occur.  In any case, Algorithm~\ref{alg:transposecache}
is coded so that every set of $W$ simultaneous}}}

\vskip-\parskip\noindent
({\sc simd}) memory accesses refers
to $W$ consecutive memory locations, so it runs much faster than a ``nontransposed'' version would;
and the result is that each thread ends up with data that other threads need.

\begin{algorithm}[t]
\caption{Compute partial sums (transposed access)}\label{alg:transposesum}
\begin{algorithmic}[1]
\CodeChunk{compute partial sums of $\theta$-$\phi$ products}
               \Bind{$c$}{$w[m][i]$}  \label{z6wordfetch}
               \ForAll{$0 \leq k < W$}    \label{z6broadcaststart}
                 \Assign{$c_{\mathrm{warp}}[k,r]$}{$c$}    \Comment{Transposed access to $c_{\mathrm{warp}}$}
               \EndFor     \label{z6broadcastend}
               \Bind{$\Var{sum}$}{$0.0$}
                 \Bind{$j$}{$0$}
                  \While{$j < (K \bmod W)$}  \label{z6remnantsumstart}
                      \Assign{$\Var{sum}$}{$\Var{sum} + (\theta_{\mathrm{local}}[j] \times \phi[c, j])$}
                      \Assign{$p[m, j]$}{$\Var{sum}$}
                      \Assign{$j$}{$j + 1$}
                 \EndWhile   \label{z6remnantsumend}
                 \While{$j < K$}
                    \For{$k$ from $0$ through $W-1$}   \label{z6firstsumprodinnerloopstart}
                       \Remark{Next line uses transposed access to $\phi$}
                       \Assign{$a[m,k]$}{$\theta_{\mathrm{local}}[j+k] \times \phi[c_{\mathrm{warp}}[r,k], j+r]$}     \label{z6phiaccess}
                    \EndFor    \label{z6firstsumprodinnerloopend}
                    \For{$k$ from $0$ through $W-1$}   \label{z6secondsumprodinnerloopstart}
                       \Remark{Next line uses transposed access to $a$}
                       \Assign{$\Var{sum}$}{$\Var{sum} + a[q \times W + k, r]$}     \label{z6aaccess}
                       \Assign{$p[m, j+k]$}{$\Var{sum}$}
                    \EndFor    \label{z6secondsumprodinnerloopend}
                    \Assign{$j$}{$j + W$}
                 \EndWhile
\End
\end{algorithmic}
\end{algorithm}

One possible remedy is to have the threads exchange data so that each thread has exactly the $\theta$ values
it needs for the rest of the computation.  Instead, we compensate for the transposition of
$\theta$ in Algorithm~\ref{alg:transposesum}.
The idea is to divide the array $\phi$ into blocks (and possibly a remnant)
and perform transposed accesses to $\phi$.  To do this, each thread needs to know what part of
the array $\phi$ every other thread is interested in; this is done through the $W \times W$ local work array $c_{\mathit{warp}}$.
In line~\ref{z6wordfetch}, each thread figures out which word is the $i$th word of its document and calls it $c$;
in lines~\hbox{\ref{z6broadcaststart}--\ref{z6broadcastend}} it then stores its value for $c$ into every element of row $r$
of the array $c_{\mathit{warp}}$.
This is not an especially fast operation, but it pays for itself later on.
The loop in lines~\hbox{\ref{z6remnantsumstart}--\ref{z6remnantsumend}} computes $\theta$-$\phi$ products and partial sums $p$
in the usual way (remember that the remnant in $\theta_{\mathit{local}}$ is not transposed), but the loop in
lines~\hbox{\ref{z6firstsumprodinnerloopstart}--\ref{z6firstsumprodinnerloopend}}
processes a block to compute product values to store into the $a$ array;
the access to $\phi$ on line~\ref{z6phiaccess} is transposed (note that the accesses to $\theta_{\mathit{local}}$
and $c_{\mathit{warp}}$ are \emph{not} transposed; because they were constructed and stored in transposed form,
normal fetches cause their values to line up correctly with the $\phi$ values obtained by a transposed fetch).
So this is pretty good; but in line~\ref{z6aaccess} we finally pay the piper: in order to have the finally computed partial sums $p$
reside in the correct thread for the binary search, it is necessary to perform a transposed access to $a$
on line~\ref{z6aaccess}; but $a$ is a local array, so transposed accesses are bad rather than good,
and this occurs in an inner loop, so performance still suffers.

\begin{algorithm}[t]
\caption{Drawing new $z$ values using a butterfly table}\label{alg:simddraw}
\begin{algorithmic}[1]
\ProcTwo{DrawZ}{$N[M], \theta[M, K], \phi[V, K]$}\ProcTwoX{$w[M][N]; \Output z[M, N]$}
%   \Remark{$M$ must be a multiple of the machine-dependent constant $W$, which is a power of 2}
   \ForAll{$0 \leq q < M/W$}
      \For{\hbox{\small\bf SIMD} $0 \leq r < W$}
         \Bind{$m$}{$q \times W + r$}
        \LocalArray{$p[K], \theta_{\mathrm{local}}[K]$}  \label{z9localdecls}
         \RegisterArray{$a[W], c_{\mathrm{warp}}[W]$}  \label{z9registerdecls}
         \UseCodeChunk{cache $\theta$ values into $\theta_{\mathrm{local}}$}
         \Bind{$i_{\mathrm{master}}$}{$0$}
          \While{$\Var{any}(i_{\mathrm{master}} < N[m])$}
            \Bind{$i$}{$\Var{min}(i_{\mathrm{master}}, N[m]-1)$}
            \UseCodeChunk{{\sc SIMD} compute butterfly partial sums}
            \Bind{$j$}{$0$}
            \UseCodeChunk{{\sc SIMD} search butterfly partial sums}
            \Assign{$z[m, i]$}{$j$}
            \Assign{$i_{\mathrm{master}}$}{$i_{\mathrm{master}} + 1$}
         \EndWhile
     \EndFor
   \EndFor
\EndProcedure
\end{algorithmic}
\end{algorithm}

\section{Butterfly-patterned Partial Sums}

We can avoid the cost of the final transposition of $a$ by not requiring the partial sums table $p$
for each thread to be entirely in the local memory of that thread.  Instead, we arrange for the
threads to ``help each other'' during the binary search, in much the same way that
each thread computes $\theta$-$\phi$ products that are actually of interest to other threads.
Moreover, we avoid computing the entire set of partial sums; instead (this is our novel contribution) we compute a
``butterfly-patterned'' table of partial sums that is sufficient to reconstruct any
needed partial sum on the fly during the binary search process.  This makes each step of
the binary search process slower, but a binary search
of a block of $W$ entries examines only $\log_2 W$ elements of the block.

Our final version is Algorithm~\ref{alg:simddraw}.  It is quite similar to Algorithm~\ref{alg:transposedraw},
but declares all local arrays in such a way as to be thread-local (and specifies that arrays $a$ and $c_{\mathit{warp}}$
are expected to reside in registers).  It uses the code chunk in Algorithm~\ref{alg:transposecache}
to cache $\theta$ values in $\theta_{\mathit{local}}$, and also uses two new code chunks:
Algorithm~\ref{alg:simdsum} creates a butterfly-patterned table of partial sums, and
Algorithm~\ref{alg:simdsearch} uses this table to perform the binary search.
For this algorithm to work properly, $W$ must be a power of 2.

\begin{algorithm}[t]
\caption{Compute a butterfly-patterned table of sums}\label{alg:simdsum}
\begin{algorithmic}[1]
\CodeChunk{{\sc SIMD} compute butterfly partial sums}
             \Bind{$c$}{$w[m][i]$}
               \ForAll{$0 \leq k < W$}
                 \Assign{$c_{\mathrm{warp}}[k]$}{$\Var{shuffle}(c,k)$}  \label{z9shuffle}
               \EndFor
               \Bind{$\Var{sum}$}{$0.0$}
                 \Bind{$j$}{$0$}
                  \While{$j < (K \bmod W)$}
                      \Assign{$\Var{sum}$}{$\Var{sum} + (\theta_{\mathrm{local}}[j] \times \phi[c, j])$}
                      \Assign{$p[j]$}{$\Var{sum}$}
                      \Assign{$j$}{$j + 1$}
                 \EndWhile
                 \While{$j < K$}
                    \For{$k$ from $0$ through $W-1$}  \label{z9unrollk}
                       \Remark{Next line uses transposed access to $\phi$}
                       \Assign{$a[k]$}{$\theta_{\mathrm{local}}[j+k] \times \phi[c_{\mathrm{warp}}[k], j+r]$}
                    \EndFor
                    \For{$b$ from $0$ through $(\log_2 W)-1$}  \label{z9unrollb}
                       \Bind{$\Var{bit}$}{$2^b$}
                       \For{$i$ from $0$ through $\frac{W}{2 \times \Var{bit}} - 1$}  \label{z9unrolli}
                          \Bind{$d$}{$2 \times \Var{bit} \times i + (\Var{bit} - 1)$}
                          \Bind{$\Var{h}$}{($\mathbf{if}\;(m \mathbin\& \Var{bit}) \neq 0$}
                          \PhantomBind{$\Var{h}$}{$\mathbf{then}\;a[d]$}
                          \PhantomBind{$\Var{h}$}{$\mathbf{else}\;a[d+\Var{bit}]$)}
                         \Bind{$v$}{$\Var{shuffleXor}(h, bit)$}  \label{z9shufflexor}
                          \If{$(r \mathbin\& \Var{bit}) \neq 0$}
                             \Assign{$a[d]$}{$a[d+\Var{bit}]$}
                          \EndIf
                          \Assign{$a[d+\Var{bit}]$}{$a[d] + v$}
                          \Assign{$p[j+d]$}{$a[d]$}
                       \EndFor
                    \EndFor
                    \Assign{$\Var{sum}$}{$\Var{sum} + a[W-1]$}
                    \Assign{$p[W-1]$}{$\Var{sum}$}
                    \Assign{$j$}{$j + W$}
                 \EndWhile
\End
\end{algorithmic}
\end{algorithm}

\newcommand\Z[3]{{\hbox{${#1}_{\ZS{#2}}^{\ZS{#3}}$}}}

\newcommand\ZS[1]{{\setbox0=\hbox{$\scriptstyle#1$}\setbox1=\hbox{$\scriptstyle9$}\relax
  \ifdim \wd0 > 1.5\wd1 \hbox{\hskip-1pt$\scriptstyle#1$}\else \hbox{$\scriptstyle#1$}\fi}}

\newcommand\ZB[3]{{\unitlength=1 in \divide\unitlength by 16
  \begin{picture}(3.3,3)
  \put(0,0){\hbox{\color[gray]{1}\vrule width 3.3\unitlength height 3\unitlength}}
  \put(1.65,1.5){\vbox to 0pt{\vss\hbox to 0pt{\hss{$\Z{#1}{#2}{#3}$}\hss}\vss}}
  \put(0,0){\line(0,1){3}}
  \put(0,0){\line(1,0){3.3}}
  \put(3.3,0){\line(0,1){3}}
  \put(0,3){\line(1,0){3.3}}
 \end{picture}}}

To save space in our figures, we introduce a special abbreviation: $\Z{m}{p}{q} = \sum a[p\mathbin{:}q] = \sum_{k=p}^{q}  \theta[m,k] \times \phi[w[m][i],k]$.
(The variable $i$ is a free parameter of this notation; when we use the notation, it will consistently refer to
a computation in one loop iteration during which $i$ is unchanging.)  The large number $m$ identifies a thread
that owns a document, and the subscript $p$ and superscript $q$ identify a subarray $a[p\mathbin{:}q]$; the symbol denotes the sum over that subarray.

For purposes of illustration, we shall (as before) assume $W=8$ and $K=19$,
so that the remnant is of size $3$ and there are two blocks of size $8$; we also assume $M=8$.

\newdimen\patternhskip \patternhskip=0.7pt
\newcommand\PH{\hskip\patternhskip}
\begin{figure}
\renewcommand\arraystretch{1.2}   
\hbox to \linewidth{\relax
\begin{tabular}{|@{$\,$}l@{\PH}l@{\PH}l@{\PH}l@{\PH}l@{\PH}l@{\PH}l@{\PH}l@{$\,$}|}
\hline
\Z{0}{0}{0}&\Z{1}{0}{0}&\Z{2}{0}{0}&\Z{3}{0}{0}&\Z{4}{0}{0}&\Z{5}{0}{0}&\Z{6}{0}{0}&\Z{7}{0}{0}\\
\Z{0}{0}{1}&\Z{1}{0}{1}&\Z{2}{0}{1}&\Z{3}{0}{1}&\Z{4}{0}{1}&\Z{5}{0}{1}&\Z{6}{0}{1}&\Z{7}{0}{1}\\
\Z{0}{0}{2}&\Z{1}{0}{2}&\Z{2}{0}{2}&\Z{3}{0}{2}&\Z{4}{0}{2}&\Z{5}{0}{2}&\Z{6}{0}{2}&\Z{7}{0}{2}\\
\hline
\Z{0}{0}{3}&\Z{1}{0}{3}&\Z{2}{0}{3}&\Z{3}{0}{3}&\Z{4}{0}{3}&\Z{5}{0}{3}&\Z{6}{0}{3}&\Z{7}{0}{3}\\
\Z{0}{0}{4}&\Z{1}{0}{4}&\Z{2}{0}{4}&\Z{3}{0}{4}&\Z{4}{0}{4}&\Z{5}{0}{4}&\Z{6}{0}{4}&\Z{7}{0}{4}\\
\Z{0}{0}{5}&\Z{1}{0}{5}&\Z{2}{0}{5}&\Z{3}{0}{5}&\Z{4}{0}{5}&\Z{5}{0}{5}&\Z{6}{0}{5}&\Z{7}{0}{5}\\
\Z{0}{0}{6}&\Z{1}{0}{6}&\Z{2}{0}{6}&\Z{3}{0}{6}&\Z{4}{0}{6}&\Z{5}{0}{6}&\Z{6}{0}{6}&\Z{7}{0}{6}\\
\Z{0}{0}{7}&\Z{1}{0}{7}&\Z{2}{0}{7}&\Z{3}{0}{7}&\Z{4}{0}{7}&\Z{5}{0}{7}&\Z{6}{0}{7}&\Z{7}{0}{7}\\
\Z{0}{0}{8}&\Z{1}{0}{8}&\Z{2}{0}{8}&\Z{3}{0}{8}&\Z{4}{0}{8}&\Z{5}{0}{8}&\Z{6}{0}{8}&\Z{7}{0}{8}\\
\Z{0}{0}{9}&\Z{1}{0}{9}&\Z{2}{0}{9}&\Z{3}{0}{9}&\Z{4}{0}{9}&\Z{5}{0}{9}&\Z{6}{0}{9}&\Z{7}{0}{9}\\
\Z{0}{0}{10}&\Z{1}{0}{10}&\Z{2}{0}{10}&\Z{3}{0}{10}&\Z{4}{0}{10}&\Z{5}{0}{10}&\Z{6}{0}{10}&\Z{7}{0}{10}\\
\hline
\Z{0}{0}{11}&\Z{1}{0}{11}&\Z{2}{0}{11}&\Z{3}{0}{11}&\Z{4}{0}{11}&\Z{5}{0}{11}&\Z{6}{0}{11}&\Z{7}{0}{11}\\
\Z{0}{0}{12}&\Z{1}{0}{12}&\Z{2}{0}{12}&\Z{3}{0}{12}&\Z{4}{0}{12}&\Z{5}{0}{12}&\Z{6}{0}{12}&\Z{7}{0}{12}\\
\Z{0}{0}{13}&\Z{1}{0}{13}&\Z{2}{0}{13}&\Z{3}{0}{13}&\Z{4}{0}{13}&\Z{5}{0}{13}&\Z{6}{0}{13}&\Z{7}{0}{13}\\
\Z{0}{0}{14}&\Z{1}{0}{14}&\Z{2}{0}{14}&\Z{3}{0}{14}&\Z{4}{0}{14}&\Z{5}{0}{14}&\Z{6}{0}{14}&\Z{7}{0}{14}\\
\Z{0}{0}{15}&\Z{1}{0}{15}&\Z{2}{0}{15}&\Z{3}{0}{15}&\Z{4}{0}{15}&\Z{5}{0}{15}&\Z{6}{0}{15}&\Z{7}{0}{15}\\
\Z{0}{0}{16}&\Z{1}{0}{16}&\Z{2}{0}{16}&\Z{3}{0}{16}&\Z{4}{0}{16}&\Z{5}{0}{16}&\Z{6}{0}{16}&\Z{7}{0}{16}\\
\Z{0}{0}{17}&\Z{1}{0}{17}&\Z{2}{0}{17}&\Z{3}{0}{17}&\Z{4}{0}{17}&\Z{5}{0}{17}&\Z{6}{0}{17}&\Z{7}{0}{17}\\
\Z{0}{0}{18}&\Z{1}{0}{18}&\Z{2}{0}{18}&\Z{3}{0}{18}&\Z{4}{0}{18}&\Z{5}{0}{18}&\Z{6}{0}{18}&\Z{7}{0}{18}\\
\hline
\end{tabular}\hfil
\begin{tabular}{|@{$\,$}l@{\PH}l@{\PH}l@{\PH}l@{\PH}l@{\PH}l@{\PH}l@{\PH}l@{$\,$}|}
\hline
\Z{0}{0}{0}&\Z{1}{0}{0}&\Z{2}{0}{0}&\Z{3}{0}{0}&\Z{4}{0}{0}&\Z{5}{0}{0}&\Z{6}{0}{0}&\Z{7}{0}{0}\\
\Z{0}{0}{1}&\Z{1}{0}{1}&\Z{2}{0}{1}&\Z{3}{0}{1}&\Z{4}{0}{1}&\Z{5}{0}{1}&\Z{6}{0}{1}&\Z{7}{0}{1}\\
\Z{0}{0}{2}&\Z{1}{0}{2}&\Z{2}{0}{2}&\Z{3}{0}{2}&\Z{4}{0}{2}&\Z{5}{0}{2}&\Z{6}{0}{2}&\Z{7}{0}{2}\\
\hline
\Z{0}{3}{3}&\Z{1}{4}{4}&\Z{0}{5}{5}&\Z{1}{6}{6}&\Z{0}{7}{7}&\Z{1}{8}{8}&\Z{0}{9}{9}&\Z{1}{10}{10}\\ 
\Z{0}{3}{4}&\Z{1}{3}{4}&\Z{2}{5}{6}&\Z{3}{5}{6}&\Z{0}{7}{8}&\Z{1}{7}{8}&\Z{2}{9}{10}&\Z{3}{9}{10}\\  
\Z{2}{3}{3}&\Z{3}{4}{4}&\Z{2}{5}{5}&\Z{3}{6}{6}&\Z{2}{7}{7}&\Z{3}{8}{8}&\Z{2}{9}{9}&\Z{3}{10}{10}\\   
\Z{0}{3}{6}&\Z{1}{3}{6}&\Z{2}{3}{6}&\Z{3}{3}{6}&\Z{4}{7}{10}&\Z{5}{7}{10}&\Z{6}{7}{10}&\Z{7}{7}{10}\\    
\Z{4}{3}{3}&\Z{5}{4}{4}&\Z{4}{5}{5}&\Z{5}{6}{6}&\Z{4}{7}{7}&\Z{5}{8}{8}&\Z{4}{9}{9}&\Z{5}{10}{10}\\     
\Z{4}{3}{4}&\Z{5}{3}{4}&\Z{6}{5}{6}&\Z{7}{5}{6}&\Z{4}{7}{8}&\Z{5}{7}{8}&\Z{6}{9}{10}&\Z{7}{9}{10}\\      
\Z{6}{3}{3}&\Z{7}{4}{4}&\Z{6}{5}{5}&\Z{7}{6}{6}&\Z{6}{7}{7}&\Z{7}{8}{8}&\Z{6}{9}{9}&\Z{7}{10}{10}\\       
\Z{0}{0}{10}&\Z{1}{0}{10}&\Z{2}{0}{10}&\Z{3}{0}{10}&\Z{4}{0}{10}&\Z{5}{0}{10}&\Z{6}{0}{10}&\Z{7}{0}{10}\\        
\hline
\Z{0}{11}{11}&\Z{1}{12}{12}&\Z{0}{13}{13}&\Z{1}{14}{14}&\Z{0}{15}{15}&\Z{1}{16}{16}&\Z{0}{17}{17}&\Z{1}{18}{18}\\ 
\Z{0}{11}{12}&\Z{1}{11}{12}&\Z{2}{13}{14}&\Z{3}{13}{14}&\Z{0}{15}{16}&\Z{1}{15}{16}&\Z{2}{17}{18}&\Z{3}{17}{18}\\  
\Z{2}{11}{11}&\Z{3}{12}{12}&\Z{2}{13}{13}&\Z{3}{14}{14}&\Z{2}{15}{15}&\Z{3}{16}{16}&\Z{2}{17}{17}&\Z{3}{18}{18}\\   
\Z{0}{11}{14}&\Z{1}{11}{14}&\Z{2}{11}{14}&\Z{3}{11}{14}&\Z{4}{15}{18}&\Z{5}{15}{18}&\Z{6}{15}{18}&\Z{7}{15}{18}\\    
\Z{4}{11}{11}&\Z{5}{12}{12}&\Z{4}{13}{13}&\Z{5}{14}{14}&\Z{4}{15}{15}&\Z{5}{16}{16}&\Z{4}{17}{17}&\Z{5}{18}{18}\\     
\Z{4}{11}{12}&\Z{5}{11}{12}&\Z{6}{13}{14}&\Z{7}{13}{14}&\Z{4}{15}{16}&\Z{5}{15}{16}&\Z{6}{17}{18}&\Z{7}{17}{18}\\      
\Z{6}{11}{11}&\Z{7}{12}{12}&\Z{6}{13}{13}&\Z{7}{14}{14}&\Z{6}{15}{15}&\Z{7}{16}{16}&\Z{6}{17}{17}&\Z{7}{18}{18}\\       
\Z{0}{0}{18}&\Z{1}{0}{18}&\Z{2}{0}{18}&\Z{3}{0}{18}&\Z{4}{0}{18}&\Z{5}{0}{18}&\Z{6}{0}{18}&\Z{7}{0}{18}\\        
\hline
\end{tabular}}
\caption{On the left, a table of partial sums for 8 threads and 19 topics; on the right,
a butterfly-patterned table of partial sums, in which each thread holds data of interest
to other threads}
\label{fig:partialsums}
\end{figure}

For each $i$, Algorithm~\ref{alg:transposedraw} uses Algorithm~\ref{alg:transposesum}
to compute, for each $m$ and $k$, $p[m,k] = \Z{m}{0}{k} $.  For our example, this produces the $p$ table shown
on the left-hand side of Figure~\ref{fig:partialsums}; each column corresponds to a warp thread
and contains all partial sums needed by that thread.
But Algorithm~\ref{alg:simddraw} uses Algorithm~\ref{alg:simdsum}
to more cheaply compute ``butterfly-patterned partial sums'' as shown
on the right-hand side of Figure~\ref{fig:partialsums}.  Not all data needed by a thread
is in the column for that thread; for example, some data needed by thread~2 appears in columns~0, 4, and~6.  Note that all the data
in the remnant and in the last line of each block \emph{is} in the column for the thread
to which it belongs---in fact, these entries are identical to those computed by Algorithm~\ref{alg:transposesum}.
The tables differ only in the first $W-1$ rows of each block.

To see why this table can be called ``butterfly-patterned,'' consider the processing of a single block.
We regard a block of $p$ as first being initialized from a block of $\theta$-$\phi$
products in the $W \times W$ local array~$a$---which, recall, is stored in transposed form.
(To simplify this part of the illustration, we will assume that the rows of the block are numbered (indexed) from
$0$ to $W-1$; it is as if the remnant has size zero and we are examining the first block.)

The ``butterfly'' portion of the algorithm sweeps over the array $p$ in a specific order,
and at each step
operates on four entries within~$p$ that are at the intersection of two rows whose indices differ by a power of~2
and two columns whose indices differ by that same power of~2.  Suppose the four values in those entries are
$\left[ \begin{array}{c|c}a & b \\\hline c & d\end{array} \right]$; they are replaced by
$\left[ \begin{array}{c|c}a & d \\\hline a+b & c+d\end{array} \right]$.
(It might seem more natural mathematically to replace
the four values with $\left[ \begin{array}{c|c}a & c \\\hline a+b & c+d\end{array} \right]$,
and that strategy also leads to a working algorithm,
but on the {\sc nvidia} Titan {\sc gpu}, at least, that computation turns out to be noticeably more expensive,
for reasons related to the precise instruction sequences required---this butterfly-patterned computation
occurs in the innermost loop, where the inclusion of even one extra instruction can significantly decrease performance.)

\newcommand\replace[4]{\hbox{$\mathcal{R}[#1,#2;#3,#4]$}}

\newcommand\diagramshrinktop{\vskip-5pt }
\newcommand\diagramshrink{\vskip-7pt }

Now, it must be admitted that if we examine the pattern in which data is transferred between rows and
columns during one of these replacement computations, we see that it is not the conventional ``$\displaystyle\Join$'' butterfly design:
\begin{center}
\vbox{\diagramshrinktop\includegraphics[width=2.0in]{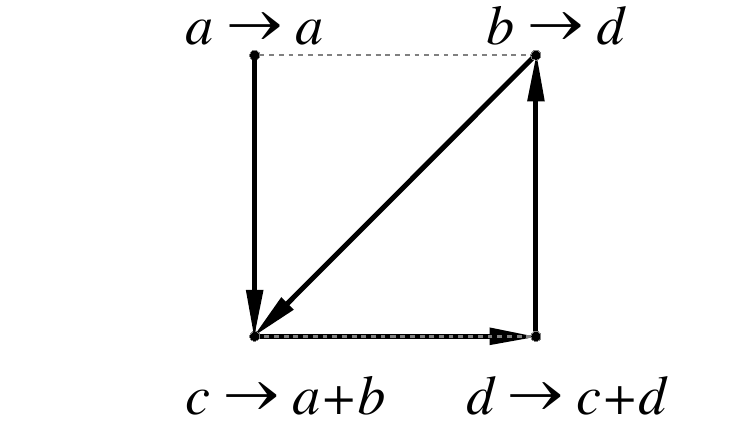}\diagramshrink}
\end{center}
To see more precisely why we use the phrase ``butterfly-patterned,'' it is helpful to consider a three-dimensional diagram
in which the vertical axis is time, the horizontal axis spans columns, and the front-to-back axis spans rows:
\begin{center}
\vbox{\diagramshrinktop\includegraphics[width=2.0in]{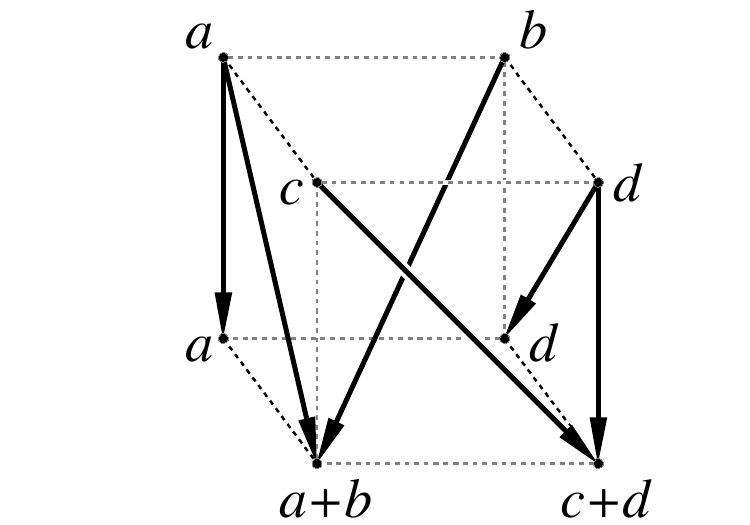}\diagramshrink}
\end{center}
The top plane of this diagram is labeled with the state of four entries before the replacement computation,
the bottom plane is labeled with the state of four entries after the replacement computation,
and the six arrows show the full pattern of transfer between the top plane and the bottom plane
(including data that remains in place, not moving in space but being carried forward in time).
The previous diagram is a vertical projection of this diagram (that is, what the 3-D diagram looks like when seen from above).

The next diagram is a front-to-back projection of this diagram (that is, what the 3-D diagram looks like when seen from the front):
\begin{center}
\vbox{\includegraphics[width=2.0in]{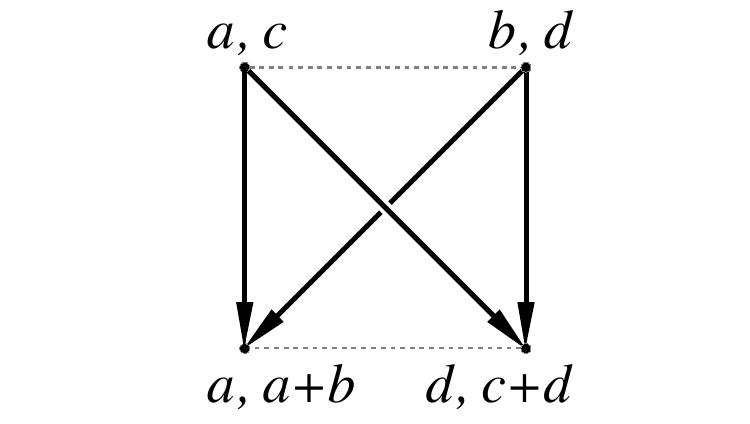}\diagramshrink}
\end{center}
and here we clearly see the standard ``$\displaystyle\Join$'' butterfly pattern as data is carried forward in time within each column and also exchanged between the two columns.  Similarly, the next diagram is a horizontal projection of the 3-D diagram (that is, what the 3-D diagram looks like when seen from the side):
\begin{center}
\vbox{\diagramshrinktop\includegraphics[width=2.0in]{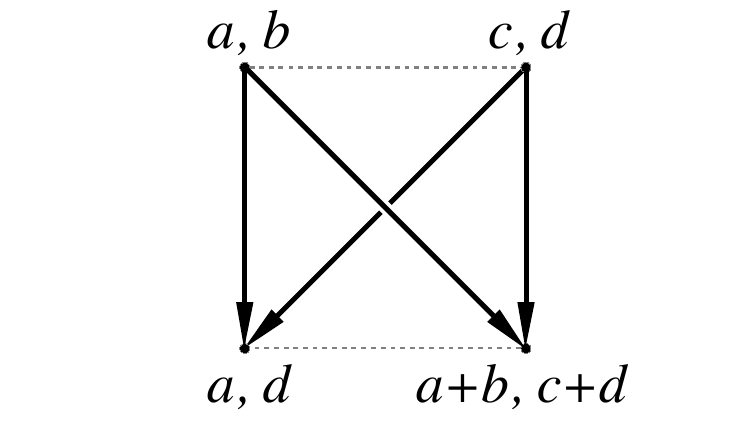}\diagramshrink}
\end{center}
and once again we clearly see the standard ``$\displaystyle\Join$'' butterfly pattern as data is carried forward in time within each row and also exchanged between the two rows.

We will use the symbol ``\replace{i}{j}{k}{l}'' to indicate application of this replacement computation to
rows $i$ and $j$ and columns $k$ and $l$----that is, to the four entries at positions
$[i,k]$, $[i,l]$, $[j,k]$, and $[j,l]$.  In our example with $W=8$, these replacement computations are performed:

\begin{center}\renewcommand\arraystretch{1.3}
\begin{tabular}{@{}llll@{}}
\hline
\replace{0}{1}{0}{1} & \replace{0}{1}{2}{3} & \replace{0}{1}{4}{5} & \replace{0}{1}{6}{7} \\
\replace{2}{3}{0}{1} & \replace{2}{3}{2}{3} & \replace{2}{3}{4}{5} & \replace{2}{3}{6}{7} \\
\replace{4}{5}{0}{1} & \replace{4}{5}{2}{3} & \replace{4}{5}{4}{5} & \replace{4}{5}{6}{7} \\
\replace{6}{7}{0}{1} & \replace{6}{7}{2}{3} & \replace{6}{7}{4}{5} & \replace{6}{7}{6}{7} \\
\hline
\replace{1}{3}{0}{2} & \replace{1}{3}{1}{3} & \replace{1}{3}{4}{6} & \replace{1}{3}{5}{7} \\
\replace{5}{7}{0}{2} & \replace{5}{7}{1}{3} & \replace{5}{7}{4}{6} & \replace{5}{7}{5}{7} \\
\hline
\replace{3}{7}{0}{4} & \replace{3}{7}{1}{5} & \replace{3}{7}{2}{6} & \replace{3}{7}{3}{7} \\
\hline
\end{tabular}
\end{center}
where replacements within a set between horizontal lines are independent
and therefore may be done sequentially or in parallel, but each set must be completed before
beginning computation of the replacements in the group below it.  Figure~\ref{fig:replacements}
shows the progress of this process with snapshots after each set of replacements is performed.
In the general case, there are $\log_2 W$ such sets of replacements.

\begin{algorithm}[t]
\caption{Searching within a butterfly-patterned table}\label{alg:simdsearch}
\begin{algorithmic}[1]
\CodeChunk{{\sc SIMD} search butterfly partial sums}
            \Bind{$u$}{random value chosen from $[0.0,1.0)$}
               \Bind{$\Var{stop}$}{$\Var{sum} \times u$}
               \Bind{$\Var{searchBase}$}{$(K \bmod W) + (W - 1)$}
               \Bind{$j$}{$0$}
               \Bind{$k$}{$\left\lfloor \frac{K}{W} \right\rfloor - 1$}
               \Remark{Binary search to find correct block of size $W$}
               \While{$j < k$}
                  \Bind{$\Var{mid}$}{$\displaystyle\left\lfloor \frac{j + k}{2} \right\rfloor$}
                  \If{$\Var{stop} < p[\Var{mid} \times W + \Var{searchBase}]$}
                     \Assign{$k$}{$\Var{mid}$}
                  \Else
                     \Assign{$j$}{$\Var{mid} + 1$}
                  \EndIf
               \EndWhile
               \Bind{$\Var{blockBase}$}{$(K \bmod W) + j \times W$}   \label{z9blockBase}
               \If{$K \geq W$}
                 \UseCodeChunk{butterfly search one block}
               \EndIf
               \If{$\Var{blockBase} > 0$}
                \If{$\Var{stop} < p[m, \Var{blockBase}-1]$}
                 \Remark{Not in a block after all, so search remnant}
                 \For{$i$ from $0$ through $(K \bmod W) - 1$}
                     \If{$\Var{stop} < p[i]$}
                        \Assign{$j$}{$i$}
                        \State {\bf break}
                     \EndIf
                  \EndFor
                \EndIf
               \EndIf
\End
\end{algorithmic}
\end{algorithm}

\begin{algorithm}[t]
\caption{Butterfly search within one $W \times W$ block}\label{alg:simdsearchoneblock}
\begin{algorithmic}[1]
\CodeChunk{butterfly search one block}
               \Bind{$\Var{lowValue}$}{($\mathbf{if}\;\Var{blockBase} > 0$}
               \PhantomBind{$\Var{lowValue}$}{$\mathbf{then}\;p[\Var{blockBase} - 1]$}
               \PhantomBind{$\Var{lowValue}$}{$\mathbf{else}\;0$)}
               \Bind{$\Var{highValue}$}{$p[\Var{blockBase} + (W - 1)]$}
               \Bind{$\Var{flip}$}{$0$}
               \Remark{Butterfly search within the block of size $W$}
               \For{$b$ from $0$ through $(\log_2 W)-1$}  \label{z9treewalkstart}
                  \Bind{$\Var{bit}$}{$2^{((\log_2 W) - 1) - b}$}
                  \Bind{$\Var{mask}$}{$((W-1) \times (2\times\Var{bit})) \mathbin\& (W-1)$}
                  \Bind{$y$}{$0$}
                  \For{$i$ from $0$ through $\frac{W}{2 \times \Var{bit}} - 1$}  \label{z9searchassistloopstart}
                     \Bind{$d$}{$(\Var{bit} - 1) + 2 \times \Var{bit} \times i$}
                     \Bind{$\Var{him}$}{$(d \mathbin\& \Var{mask})+(r \mathbin\& \mathord\neg \Var{mask})$}
                     \Bind{$\Var{hisBlockBase}$}{$\Var{shuffle}(\Var{blockBase}, \Var{him})$}
                     \Bind{$t$}{$\Var{shuffleXor}(p[\Var{hisBlockBase} + d], \Var{flip})$}  \label{z9searchassist}
                     \If{$((r \oplus d) \mathbin\& \Var{mask}) = 0$}
                        \Assign{$y$}{$t$}
                     \EndIf
                  \EndFor  \label{z9searchassistloopend}
                  \Bind{$\Var{compareValue}$}{($\mathbf{if}\;(r \mathbin\& \Var{bit}) \neq 0$}
                  \PhantomBind{$\Var{compareValue}$}{$\mathbf{then}\;\Var{highValue} - y$}
                  \PhantomBind{$\Var{compareValue}$}{$\mathbf{else}\;\Var{lowValue} + y$)}
                  \If{$stop < \Var{compareValue}$}
                     \Assign{$\Var{highValue}$}{$\Var{compareValue}$}
                     \Assign{$\Var{flip}$}{$\Var{flip} \oplus (\Var{bit} \mathbin\& r)$}
                  \Else
                     \Assign{$\Var{lowValue}$}{$\Var{compareValue}$}
                     \Assign{$\Var{flip}$}{$\Var{flip} \oplus (\Var{bit} \mathbin\& \mathord\neg r)$}
                  \EndIf
               \EndFor  \label{z9treewalkend}
               \Assign{$j$}{$\Var{blockBase} + (\Var{flip} \oplus r)$}
\End
\end{algorithmic}
\end{algorithm}

After all replacements have been done on a block of size $W \times W$,
the entry in row $i$ and column $j$ contains the value $\Z{u}{v}{w}$ where
$m = i \oplus (i+1)$,
$k = \left\lfloor \frac{m}{2} \right\rfloor$,
$u = (i \mathbin\& \mathord\neg m) + (j \mathbin\& m)$,
$v = j \mathbin\& (\mathord\neg k)$, and $w =v + k$.
We use the symbol ``$\mathord\neg$'' to indicate the bitwise negation (that is, {\sc not}) of an integer
represented in binary form; we also use the symbol ``$\mathbin\&$'' to indicate the
the bitwise conjunction (that is, {\sc and}) of two binary integers, and
the symbol ``$\oplus$'' to indicate
the bitwise exclusive {\sc or} (that is, {\sc xor}) of two binary integers.

Algorithm~\ref{alg:simdsum} performs this computation.  The function $\Var{shuffle}$ is used
in line~\ref{z9shuffle} to broadcast values from each thread of a warp to all the others,
and the function $\Var{shuffleXor}$ is used
in line~\ref{z9shufflexor} to exchange values between pairs of threads whose thread numbers
differ by a power of~2.  These are precisely the {\sc cuda} intrinsic functions
{\tt \lower0.9ex\hbox{-}\lower0.9ex\hbox{-}shfl} and {\tt \lower0.9ex\hbox{-}\lower0.9ex\hbox{-}shfl\lower0.9ex\hbox{-}xor}
\cite{CUDA-handbook,CUDA-website-shuffle-functions}.

% (dotimes (i 8)
%   (dotimes (j 8)
%     (let* ((m (logxor i (+ i 1)))
%     (u (logior (logand i (lognot m)) (logand j m)))
%     (k (ash m -1))
%     (v (logand j (lognot k)))
%     (w (logior j k)))
%      (princ (format "\\Z{%d}{%d}{%d}&" u v w))))
%   (terpri))

Within a butterfly-patterned block of partial sums,
Algorithm~\ref{alg:simdsearch} performs a binary search as follows.
The $\Var{stop}$ value is computed exactly as in Algorithm~\ref{alg:basicdraw}, and a block to be searched is identified
by performing a binary search on the subarray consisting of just the last row of each block;
this identifies a specific block to search.  If $K \geq W$, then there is at least one block,
and it is searched, but it is possible that the desired $\Var{stop}$ value does not lie within that
block; in that case, the remnant is searched using a linear search.

\begin{figure*}
\renewcommand\arraystretch{1.2} \patternhskip=3pt   
\hbox to \textwidth{\relax
\begin{tabular}{|@{$\,$}l@{\PH}l@{\PH}l@{\PH}l@{\PH}l@{\PH}l@{\PH}l@{\PH}l@{$\,$}|}
\hline
\Z{0}{0}{0}&\Z{0}{1}{1}&\Z{0}{2}{2}&\Z{0}{3}{3}&\Z{0}{4}{4}&\Z{0}{5}{5}&\Z{0}{6}{6}&\Z{0}{7}{7}\\
\Z{1}{0}{0}&\Z{1}{1}{1}&\Z{1}{2}{2}&\Z{1}{3}{3}&\Z{1}{4}{4}&\Z{1}{5}{5}&\Z{1}{6}{6}&\Z{1}{7}{7}\\
\Z{2}{0}{0}&\Z{2}{1}{1}&\Z{2}{2}{2}&\Z{2}{3}{3}&\Z{2}{4}{4}&\Z{2}{5}{5}&\Z{2}{6}{6}&\Z{2}{7}{7}\\
\Z{3}{0}{0}&\Z{3}{1}{1}&\Z{3}{2}{2}&\Z{3}{3}{3}&\Z{3}{4}{4}&\Z{3}{5}{5}&\Z{3}{6}{6}&\Z{3}{7}{7}\\
\Z{4}{0}{0}&\Z{4}{1}{1}&\Z{4}{2}{2}&\Z{4}{3}{3}&\Z{4}{4}{4}&\Z{4}{5}{5}&\Z{4}{6}{6}&\Z{4}{7}{7}\\
\Z{5}{0}{0}&\Z{5}{1}{1}&\Z{5}{2}{2}&\Z{5}{3}{3}&\Z{5}{4}{4}&\Z{5}{5}{5}&\Z{5}{6}{6}&\Z{5}{7}{7}\\
\Z{6}{0}{0}&\Z{6}{1}{1}&\Z{6}{2}{2}&\Z{6}{3}{3}&\Z{6}{4}{4}&\Z{6}{5}{5}&\Z{6}{6}{6}&\Z{6}{7}{7}\\
\Z{7}{0}{0}&\Z{7}{1}{1}&\Z{7}{2}{2}&\Z{7}{3}{3}&\Z{7}{4}{4}&\Z{7}{5}{5}&\Z{7}{6}{6}&\Z{7}{7}{7}\\
\hline
\multicolumn{8}{c}{\small Transposed $\theta$-$\phi$ products}
\end{tabular}\hfil
\begin{tabular}{|@{$\,$}l@{\PH}l@{\PH}l@{\PH}l@{\PH}l@{\PH}l@{\PH}l@{\PH}l@{$\,$}|}
\hline
\Z{0}{0}{0}&\Z{1}{1}{1}&\Z{0}{2}{2}&\Z{1}{3}{3}&\Z{0}{4}{4}&\Z{1}{5}{5}&\Z{0}{6}{6}&\Z{1}{7}{7}\\
\Z{0}{0}{1}&\Z{1}{0}{1}&\Z{0}{2}{3}&\Z{1}{2}{3}&\Z{0}{4}{5}&\Z{1}{4}{5}&\Z{0}{6}{7}&\Z{1}{6}{7}\\
\Z{2}{0}{0}&\Z{3}{1}{1}&\Z{2}{2}{2}&\Z{3}{3}{3}&\Z{2}{4}{4}&\Z{3}{5}{5}&\Z{2}{6}{6}&\Z{3}{7}{7}\\
\Z{2}{0}{1}&\Z{3}{0}{1}&\Z{2}{2}{3}&\Z{3}{2}{3}&\Z{2}{4}{5}&\Z{3}{4}{5}&\Z{2}{6}{7}&\Z{3}{6}{7}\\
\Z{4}{0}{0}&\Z{5}{1}{1}&\Z{4}{2}{2}&\Z{5}{3}{3}&\Z{4}{4}{4}&\Z{5}{5}{5}&\Z{4}{6}{6}&\Z{5}{7}{7}\\
\Z{4}{0}{1}&\Z{5}{0}{1}&\Z{4}{2}{3}&\Z{5}{2}{3}&\Z{4}{4}{5}&\Z{5}{4}{5}&\Z{4}{6}{7}&\Z{5}{6}{7}\\
\Z{6}{0}{0}&\Z{7}{1}{1}&\Z{6}{2}{2}&\Z{7}{3}{3}&\Z{6}{4}{4}&\Z{7}{5}{5}&\Z{6}{6}{6}&\Z{7}{7}{7}\\
\Z{6}{0}{1}&\Z{7}{0}{1}&\Z{6}{2}{3}&\Z{7}{2}{3}&\Z{6}{4}{5}&\Z{7}{4}{5}&\Z{6}{6}{7}&\Z{7}{6}{7}\\
\hline
\multicolumn{8}{c}{\small After first set}
\end{tabular}\hfil
\begin{tabular}{|@{$\,$}l@{\PH}l@{\PH}l@{\PH}l@{\PH}l@{\PH}l@{\PH}l@{\PH}l@{$\,$}|}
\hline
\Z{0}{0}{0}&\Z{1}{1}{1}&\Z{0}{2}{2}&\Z{1}{3}{3}&\Z{0}{4}{4}&\Z{1}{5}{5}&\Z{0}{6}{6}&\Z{1}{7}{7}\\
\Z{0}{0}{1}&\Z{1}{0}{1}&\Z{2}{2}{3}&\Z{3}{2}{3}&\Z{0}{4}{5}&\Z{1}{4}{5}&\Z{2}{6}{7}&\Z{3}{6}{7}\\
\Z{2}{0}{0}&\Z{3}{1}{1}&\Z{2}{2}{2}&\Z{3}{3}{3}&\Z{2}{4}{4}&\Z{3}{5}{5}&\Z{2}{6}{6}&\Z{3}{7}{7}\\
\Z{0}{0}{3}&\Z{1}{0}{3}&\Z{2}{0}{3}&\Z{3}{0}{3}&\Z{0}{4}{7}&\Z{1}{4}{7}&\Z{2}{4}{7}&\Z{3}{4}{7}\\
\Z{4}{0}{0}&\Z{5}{1}{1}&\Z{4}{2}{2}&\Z{5}{3}{3}&\Z{4}{4}{4}&\Z{5}{5}{5}&\Z{4}{6}{6}&\Z{5}{7}{7}\\
\Z{4}{0}{1}&\Z{5}{0}{1}&\Z{6}{2}{3}&\Z{7}{2}{3}&\Z{4}{4}{5}&\Z{5}{4}{5}&\Z{6}{6}{7}&\Z{7}{6}{7}\\
\Z{6}{0}{0}&\Z{7}{1}{1}&\Z{6}{2}{2}&\Z{7}{3}{3}&\Z{6}{4}{4}&\Z{7}{5}{5}&\Z{6}{6}{6}&\Z{7}{7}{7}\\
\Z{4}{0}{3}&\Z{5}{0}{3}&\Z{6}{0}{3}&\Z{7}{0}{3}&\Z{4}{4}{7}&\Z{5}{4}{7}&\Z{6}{4}{7}&\Z{7}{4}{7}\\
\hline
\multicolumn{8}{c}{\small After second set}
\end{tabular}\hfil
\begin{tabular}{|@{$\,$}l@{\PH}l@{\PH}l@{\PH}l@{\PH}l@{\PH}l@{\PH}l@{\PH}l@{$\,$}|}
\hline
\Z{0}{0}{0}&\Z{1}{1}{1}&\Z{0}{2}{2}&\Z{1}{3}{3}&\Z{0}{4}{4}&\Z{1}{5}{5}&\Z{0}{6}{6}&\Z{1}{7}{7}\\
\Z{0}{0}{1}&\Z{1}{0}{1}&\Z{2}{2}{3}&\Z{3}{2}{3}&\Z{0}{4}{5}&\Z{1}{4}{5}&\Z{2}{6}{7}&\Z{3}{6}{7}\\
\Z{2}{0}{0}&\Z{3}{1}{1}&\Z{2}{2}{2}&\Z{3}{3}{3}&\Z{2}{4}{4}&\Z{3}{5}{5}&\Z{2}{6}{6}&\Z{3}{7}{7}\\
\Z{0}{0}{3}&\Z{1}{0}{3}&\Z{2}{0}{3}&\Z{3}{0}{3}&\Z{4}{4}{7}&\Z{5}{4}{7}&\Z{6}{4}{7}&\Z{7}{4}{7}\\
\Z{4}{0}{0}&\Z{5}{1}{1}&\Z{4}{2}{2}&\Z{5}{3}{3}&\Z{4}{4}{4}&\Z{5}{5}{5}&\Z{4}{6}{6}&\Z{5}{7}{7}\\
\Z{4}{0}{1}&\Z{5}{0}{1}&\Z{6}{2}{3}&\Z{7}{2}{3}&\Z{4}{4}{5}&\Z{5}{4}{5}&\Z{6}{6}{7}&\Z{7}{6}{7}\\
\Z{6}{0}{0}&\Z{7}{1}{1}&\Z{6}{2}{2}&\Z{7}{3}{3}&\Z{6}{4}{4}&\Z{7}{5}{5}&\Z{6}{6}{6}&\Z{7}{7}{7}\\
\Z{0}{0}{7}&\Z{1}{0}{7}&\Z{2}{0}{7}&\Z{3}{0}{7}&\Z{4}{0}{7}&\Z{5}{0}{7}&\Z{6}{0}{7}&\Z{7}{0}{7}\\
\hline
\multicolumn{8}{c}{\small After third set}
\end{tabular}}
\caption{Generation of butterfly-patterned partial sums for a $W \times W$ block in three steps, each using a set of four-element replacement computations}
\label{fig:replacements}
\end{figure*}

In order to search within a block, Algorithm~\ref{alg:simdsearchoneblock} maintains
two additional state variables $\Var{lowValue}$
and $\Var{highValue}$.  An invariant is that if thread $m$ has
indices $j$ through $k$ of a block still under consideration, then $\Var{lowValue} = \Z{m}{0}{\Var{blockBase}+j-1}$
and $\Var{highValue} = \Z{m}{0}{\Var{blockBase}+k}$.  In order to cut the search range in half, the binary search needs to
compare the $\Var{stop}$ value to the midpoint value $\Z{m}{0}{\Var{blockBase}+\Var{mid}}$
where $\Var{mid}=\left\lfloor \frac{j+k}{2} \right\rfloor$; in Algorithm~\ref{alg:basicbinarysearch}
this value is of course an entry in the $p$ array, namely $p[m, \Var{blockBase}+\Var{mid}]$,
but in Algorithm~\ref{alg:simdsearchoneblock} the midpoint value is \emph{calculated} by choosing an appropriate entry from
the butterfly-patterned $p$~array and then either adding it to $\Var{lowValue}$ or subtracting it from $\Var{highValue}$.
Whether to add or subtract on iteration number $b$ (where the $\log_2 W$ iterations are numbered starting from 0)
depends on whether bit $b$ (counting from the right starting at 0) of the binary representation
of $m$ is $0$ or $1$, respectively.  Depending on the result of the comparison of the midpoint value
with the $\Var{stop}$ value,
the midpoint value is assigned to either $\Var{lowValue}$ and $\Var{highValue}$, maintaining the invariant.
Also depending on the result of the comparison of the midpoint value
with the $\Var{stop}$ value, a bit of a third state variable $\Var{flip}$ (initially $0$) is updated.
When the binary search is complete, the correct index to select is computed from the value in $\Var{flip}$.

A normal binary search effectively walks down a binary decision tree,
where each node of the tree is labeled with a (normal) partial sum; but Algorithm~\ref{alg:simdsearchoneblock} walks
down a tree that is labeled with entries from the butterfly-patterned table of partial sums.
For example, referring to the first seven rows of the upper block in the right-hand diagram in Figure~\ref{fig:partialsums},
the entries relevant to thread 5 form this tree:

\newcommand\centerit[1]{\vbox to 0pt{\vss\hbox to 0pt{\hss#1\hss}\vss}}

\vskip 1pt
\hbox to \linewidth{\relax
\hsize=0.43\linewidth
\vbox{\unitlength=1 in \divide\unitlength by 16\noindent
\begin{picture}(24,18)(1.5,0)
\put(6,9){\line(1,1){6}}
\put(18,9){\line(-1,1){6}}
\put(3,3){\line(1,2){3}}
\put(9,3){\line(-1,2){3}}
\put(15,3){\line(1,2){3}}
\put(21,3){\line(-1,2){3}}
\put(12,15){\centerit{\ZB{5}{7}{10}}}
\put(6,9){\centerit{\ZB{5}{3}{4}}}
\put(18,9){\centerit{\ZB{5}{7}{8}}}
\put(3,3){\centerit{\ZB{5}{4}{4}}}
\put(9,3){\centerit{\ZB{5}{6}{6}}}
\put(15,3){\centerit{\ZB{5}{8}{8}}}
\put(21,3){\centerit{\ZB{5}{10}{10}}}
\end{picture}
}\hfil
{\hsize=0.57\linewidth
\vbox{\noindent\strut
Starting with $\Var{lowValue} = \Z{5}{0}{2}$ and $\Var{highValue} = \Z{5}{0}{10}$,
lines~\hbox{\ref{z9treewalkstart}--\ref{z9treewalkend}} of Algorithm~\ref{alg:simdsearchoneblock}
can walk down the tree, at each node calculating a new midpoint
by adding the node label to $\Var{lowValue}$ or subtracting it from $\Var{highValue}$ as appropriate.\strut}}}

\vskip-\parskip
The threads in a warp assist one another in fetching these tree nodes
using the loop in lines~\hbox{\ref{z9searchassistloopstart}--\ref{z9searchassistloopend}};
the function $\Var{shuffleXor}$ effects this data transfer in line \ref{z9searchassist}.

\begin{figure}
\includegraphics[width=\linewidth]{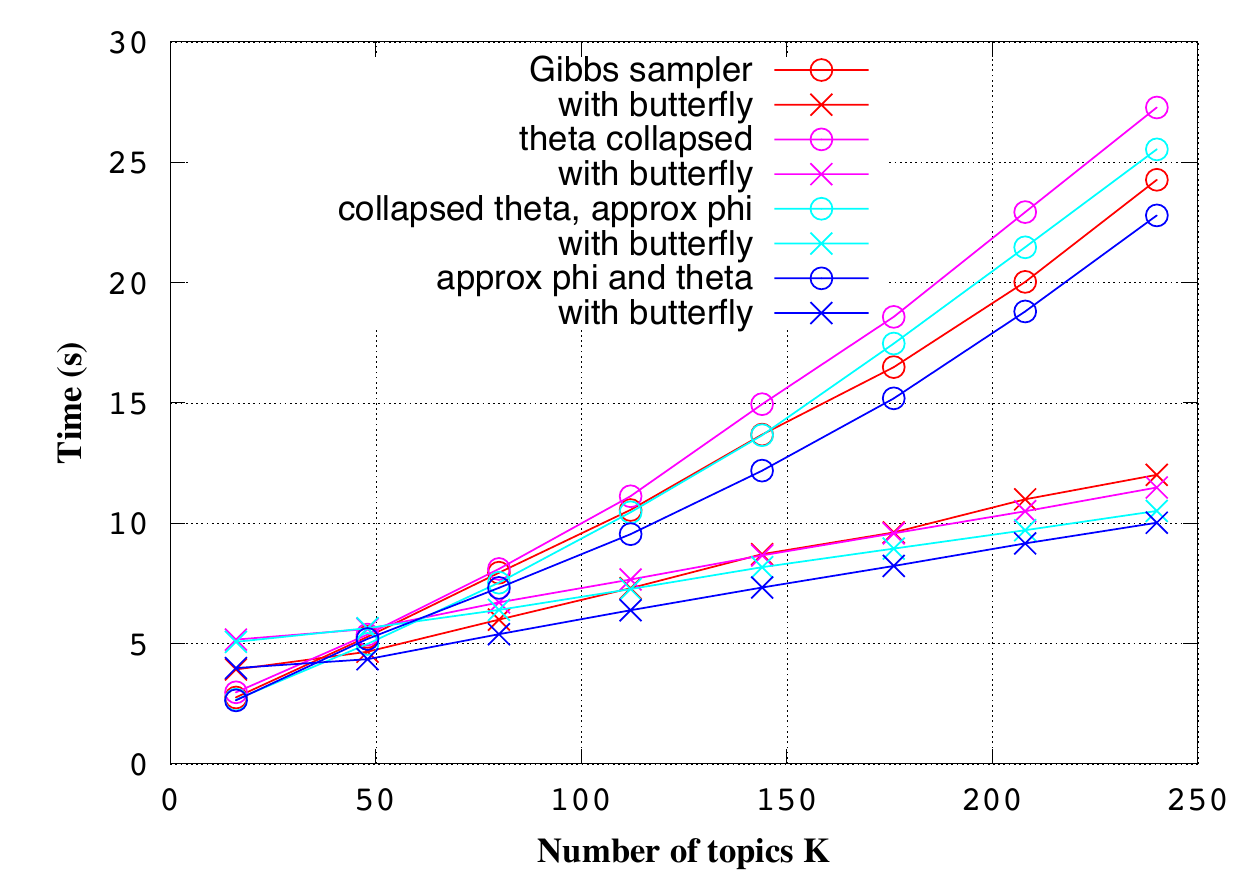}
\caption{Execution time ($K\mathbin=32k+16, 0\mathbin\leq k\mathbin\leq 7$)}\label{fig:performance}
\end{figure}

\section{Evaluation}

We coded four versions of a complete {\sc lda} Gibbs-sampler topic-modeling algorithm in {\sc cuda} for an {\sc nvidia}
Titan Black {\sc gpu} ($W\mathbin=\,32$).  For each version we tested two variants, one using Algorithm~\ref{alg:basicdraw} (using the binary search
of Algorithm~\ref{alg:basicbinarysearch}) and one using Algorithm~\ref{alg:simddraw}.  All eight variants
were tested for speed using a Wikipedia-based
dataset with number of documents $M=43556$, vocabulary size $V=37286$, total number of words in corpus $\Sigma N=3072662$
(therefore average document size $(\Sigma N)/M \approx 70.5$),
and maximum document size $\max N=307$.  Each variant was measured using eight different values for the number of topics $K$ (16, 48, 80, 112, 144, 176, 208, and 240),
in each case performing 100 sampling iterations and measuring the execution time of the entire application, not just the
part that draws $z$ values.  Best performance requires unrolling three loops in Algorithm~\ref{alg:simdsum};
we had to manually unroll the loop that starts on line~\ref{z9unrollb}, and the {\sc cuda} compiler then automatically
unrolled the loops that start on lines~\ref{z9unrollk} and~\ref{z9unrolli}.
The performance results are shown in Figure~\ref{fig:performance}.
The butterfly variants are faster for $K \geq 80$.  For $K \geq 200$,
for each of the four versions the butterfly variant is more than twice as fast.

\section{Related Work}

Because the computed probabilities are relative in our LDA application, it is necessary to compute all of them and then to compute,
if nothing else, their sum, so that the relative probabilities can be effectively normalized.  Therefore every method for
drawing from a discrete distribution represented by a set of relative probabilities
involves some amount of preprocessing before drawing from the distribution.
The various algorithms in the literature have differing tradeoffs according to what technique is used for preprocessing
what technique is used for drawing; some algorithms also accommodate incremental updating of the relative probabilities
by providing a technique for incremental preprocessing.

Instead of doing a binary search on the table of partial sums, one can instead (as Marsaglia~\cite{Marsaglia-1963} observes in passing)
construct a search tree using the principles of Huffman encoding~\cite{Huffman-1952} (independently rediscovered
by Zimmerman\cite{Zimmerman-1959}) to minimize the expected number of comparisons.  In either case
the complexity of the search is $O(\log n)$, but the optimized search may have a smaller constant,
obtained at the expense of a preprocessing step that must sort the relative probabilities and
therefore have complexity $\Omega(n \log n)$.

Walker~\cite{Walker-arbitrary-1974,Walker-TOMS-1977} describes what is now known as the ``alias'' method,
in which $n$ relative probabilities are preprocessed into two additional tables $F$ and $A$ of length $n$.  To draw a value
from the distribution, let $k$ be a integer chosen uniformly at random from $\{ 0, 1, 2, \ldots , n-1\}$
and let $u$ be chosen uniformly at random from the real interval $[0,1)$.  Then the value drawn from the
distribution is
$$ \mathbf{if}\;u < F_k \;\mathbf{then}\; k \;\mathbf{else}\;A_k $$
Therefore, once the tables $F$ and $A$ have been produced, the complexity of drawing a value from the distribution is $O(1)$,
assuming that the cost of an array access is $O(1)$.
Walker's method~\cite{Walker-TOMS-1977} for producing the tables $F$ and $A$ requires time $\Theta(n^2)$;
it is easy to reduce this to $\Omega(n \log n)$ by sorting the probabilities~\cite[exercise 3.4.1-7]{KNUTH-VOLUME-2}
and then using, say, priority heaps instead of a list for the intermediate data structure.
Either version heuristically attempts to minimize the probability of having to access the table $A$.

Vose~\cite{Vose-1991} describes a preprocessing algorithm, with proof, that further reduces the preprocessing complexity
of the alias method to $\Theta(n)$.  The tradeoff that permits this improvement is that the preprocessing algorithm makes
no attempt to minimize the probability of accessing the array $A$.

Matias~{\it et~al.}~\cite{Matias-1993} describe a technique for preprocessing a set of relative probabilities into a set of trees,
after which a sequence of intermixed generate (draw) and update operations can be performed, where an update operation changes
just one of the relative probabilities; a single generate operation takes $O(\log^* n)$ expected time,
and a single update operation takes $O(\log^* n)$ amortized expected time.

Li~{\it et~al.}~\cite{li_reducing_14} describe a modified LDA topic modeling algorithm, which they call
Metropolis-Hastings-Walker sampling, that uses Walker's alias method but amortizes the cost of constructing
the table by drawing from the same table during multiple consecutive sampling iterations of a Metropolis-Hastings sampler;
their paper provides some justification for why it is acceptable to use a ``slightly stale'' alias table (their words)
for the purposes of this application.

\section{Conclusions}

The technique of constructing butterfly-patterned partial sums appears to be best suited
for situations where a SIMD processor is used to compute tables of relative probabilities for multiple discrete distributions,
each of which is then used just once to draw a single value, and where each thread, when computing its table,
must fetch data from a contiguous region of memory whose address is computed from other data.
The LDA application for which we developed the technique has these characteristics.
The technique uses transposed memory access in order to allow a SIMD memory controller
to touch only one or two cache lines on each fetch, then cheaply constructs a set of partial sums
that are just adequate to allow partial sums actually needed to be constructed on the fly
during the course of a binary search.
For a complete {\sc lda} Gibbs-sampler topic-modeling algorithm coded in {\sc cuda} for an {\sc nvidia}
Titan Black {\sc gpu} and already tuned as best we could for high performance, the butterfly-patterned
partial-sums technique further improves the speed of the overall application by at least a factor of 2
when the number of topics is greater than 200.


\bibliographystyle{plainurl}
\bibliography{butterfly}

\end{document}